\documentclass[journal,comsoc]{IEEEtran}
\usepackage[T1]{fontenc}
\usepackage{placeins}
\ifCLASSINFOpdf
\else
\fi
\usepackage{graphicx}
\graphicspath{ {./images/} }
\usepackage{threeparttable}
\usepackage[sorting = none,backend=bibtex,style=IEEEtran, style=numeric-comp]{biblatex}

\bibliography{reference.bib}

\usepackage{amsmath}
\usepackage[cmintegrals]{newtxmath}

\begin{document}

\title{Recent Trends in Food Intake Monitoring using Wearable Sensors}

\author{Muhammad~Usman and~Huanhuan~Chen,~\IEEEmembership{Senior~Member,~IEEE}
\thanks{M. Usman and H. Chen (*corresponding authors) are with the School of Computer Science and Technology, University of Science and Technology of China, Hefei 230027, China  e-mail: muhammadusman@mail.ustc.edu.cn, hchen@ustc.edu.cn}
}

\markboth{IEEE Communications Surveys \& Tutorials ,~Vol.**, No.**, ***.~2021}%
{Shell \MakeLowercase{\textit{et al.}}: Bare Demo of IEEEtran.cls for IEEE Communications Society Journals}

\maketitle

\begin{abstract}
Obesity and being over-weight add to the risk of some major life threatening diseases including heart failure, cancer and diabetes. According to W.H.O., a considerable population suffers from these disease whereas poor nutrition plays an important role in this context. Nutrition monitoring becomes critical in these circumstances that has been targeted by researchers in the past leading to the development of food activity monitoring systems. Traditional systems like Food Diaries allow manual record keeping of eating activities over time, and conduct nutrition analysis. However, these systems are prone to the problems of manual record keeping and biased-reporting. Therefore, recently, the research community has focused on designing automatic food monitoring systems since the last decade which consist of one or multiple wearable sensors. These systems aim at providing different macro and micro activity detections like chewing, swallowing, eating episodes, and food types as well as estimations like food mass and eating duration. Researchers have emphasized on high detection accuracy, low estimation errors, un-intrusive nature, low cost and real life implementation while designing these systems. Although a considerable amount of research work has been done in this domain, a comprehensive automatic food monitoring system has yet not been developed. Moreover, according to the best of our knowledge, there is no comprehensive survey in this field that delineates the automatic food monitoring paradigm, covers a handful number of research studies, analyses these studies against food intake monitoring tasks using various parameters, enlists the limitations and sets up future directions. In this research work, we delineate the automatic food intake monitoring paradigm and present a survey of research studies. With special focus on studies with wearable sensors, we analyze these studies against food activity monitoring tasks. We provide brief comparison of these studies along with shortcomings based upon experimentation results conducted under these studies. We setup future directions at the end to facilitate the researchers working in this domain.
\end{abstract}


\IEEEpeerreviewmaketitle

\section{Introduction}

\IEEEPARstart{O} {BESITY} and being over-weight can increase the risk of different serious diseases including heart diseases, diabetes and cancer. According to World Health Organization (WHO), there were 13\% obese adults and 39\% overweight adults worldwide in 2019. Moreover, 10 countries in the world have 45\% or higher population suffering from obesity \cite{who2020,procon2020}. Unhealthy lifestyles and diet are the main contributors to obesity and overweight. Due to the severe effects of obesity and over-weight on human health, the preventative healthcare includes methods to control these, mainly via regular exercise and a balanced diet. In order to observe the daily nutrition, healthcare practitioners rely heavily on the observations provided by individuals. These observations cannot be trusted for accuracy due to human memory issues and inability to manually monitor the nutrition over a period of time on regular basis. Researchers in the past worked out on food monitoring systems to assist the healthcare practitioners, particularly nutritionists in order to make decisions on individual’s diet plans.

Food intake monitoring systems can majorly be divided into two categories. Traditional systems allow manually record keeping of the eating activities over time, and allow nutrition analysis. These systems rely heavily on self-reporting, and manual recording of eating activities\cite{bedri2017earbit,muhlheim1998unsuccessful,doulah2017meal}. The other category, termed as automated systems, consists of techniques which automatically monitor eating behavior with minimum or no interaction of individuals. These approaches rely heavily on wearable sensors which track human body activities when placed at different locations on the body, and therefore have the ability to cover pitfall of the traditional systems, for example: \cite{bedri2017earbit,bi2018auracle,doulah2017meal,farooq2016detection,mirtchouk2017recognizing}. Modern studies have focused the second category encouraged by the success of initial experiments conducted by the early researchers in this area. 

Automatic food intake monitoring systems utilize wearable or non-wearable sensors targeting goals including eating detection, chewing and swallowing detection, eating episode detection, food type detection, eating duration estimation and food mass estimation. These systems can be categorized into uni-sensor and multi-sensor approaches. Inertial sensors, microphone sensors and proximity sensors are majorly used in uni-sensor based approaches (for example;\cite{stankoski2020real,farooq2016automatic,gao2016ihear}), whereas combination of two or more sensors is deployed in multi-sensor based approaches (for example; \cite{zhang2016bite,farooq2016novel,yang2019statistical}).

The automatic food intake monitoring systems are around since a decade, whereas researchers have emphasized on detection accuracy, low estimation errors, un-intrusive nature, scalability, usability in real world environment and implementation with low cost. While there have been successful implementations of some methodologies with good detection accuracy, low estimation errors and applicability in different living conditions, still there is no approach which truly satisfies all of the aforementioned properties. Previously, some surveys have been published in this domain, however these surveys are either specific to a single activity detection, do not cover a handful number of studies or do not analyze all characteristics mentioned above \cite{kalantarian2017survey,selamat2020automatic}. According to the best of our knowledge, there is no comprehensive survey in this area, which delineates the food monitoring system paradigm, covers a handful number of research studies, analyses these studies with different parameters and sets up future directions. We conduct this survey to meet these challenges to assist future studies in this area. Specifically, in this survey paper, we make the following contributions: 

1)	\textit{Definition}. We delineate the framework of automatic food monitoring systems with all components, data flow and work flow. We also provide overview of basic concepts, terminologies, definitions, algorithms and evaluation methods used in food monitoring systems.

2)	\textit{Survey}. We provide a comprehensive survey of 64 research papers across different publishers in automatic food monitoring domain between 2015 and 2020, with special focus on approaches with wearable sensors. 

3)	\textit{Analysis}. We conduct a critical analysis of all research papers by describing the motivation, contribution, experimentation and limitations against each manuscript. We also compare the results of the research studies under automatic food monitoring system tasks including eating activity detection, chewing and swallowing detection, food type detection and eating episode detection, and different estimations.

4)	\textit{Future Directions}. In this paper we identify the key challenges and open problems in the research area, and enlist potential guidelines for future research. 

Rest of the paper is organized as follows; Section II provides information on the basic concepts and terminologies used by researchers while reporting food monitoring systems. Section III is dedicated to provide reviews of the major techniques in food monitoring systems. In Section IV, we provide critical analysis of these approaches, along with setting up the future guidelines. At the end we conclude this research work in conclusion section. Appendix A contains the list of abbreviations used in the aforementioned sections.

\section{Concepts and terminologies in food intake monitoring systems}
 \begin{figure*}[ht]
 \centering

 \includegraphics{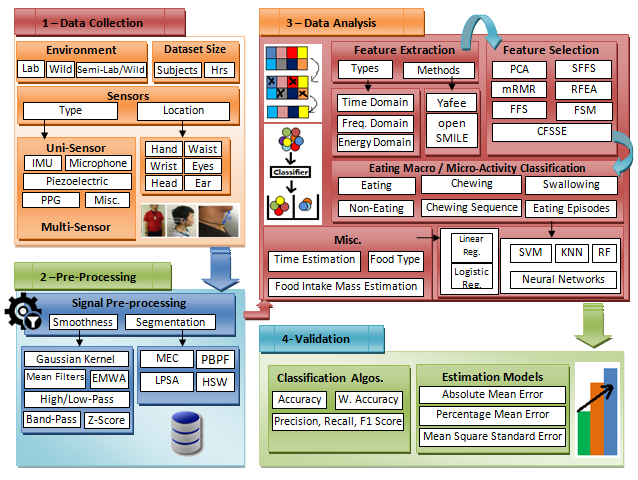}
 \caption{The generalized framework of automatic food activity monitoring systems}
 \label{fig:framework} 
 \end{figure*}
In this section, we define concepts and terminologies related to automatic food monitoring systems. As depicted in Fig. \ref{fig:framework}, an automatic food monitoring system involves four basic modules include data collection, pre-processing, data analysis and evaluation modules. 

In automatic food monitoring systems, the process generally starts with the data collection phase with the aim to save sensor data temporarily or permanently for further processing. This phase includes the procedures to collect data required for the analysis phase. Environment refers to the data collection surroundings, for example laboratory, natural (wild), semi-controlled lab and semi-free living conditions etc.

Dataset is normally characterized by the number of subjects and the number of hours of recording. Some researchers also provide information on the health conditions and gender. During data collection, single or multiple sensors have been utilized at different locations on the body. Major sensors include IMU, Piezoelectric, and microphone sensors. The pre-processing phase aims to clean the data by removing noise and apply smoothness techniques. Major smoothness techniques include high/low pass filters, z-score, and band-pass filters. Pre-processing also segments the sensor data to create window frames of a particular length. Segmentation is done using techniques like EMWA, MEC or PBPF etc. The analysis component provides feature extraction and selection ability. Normally time, frequency and energy domain features are extracted using different methods. Not all the features play role in classification, so authors normally apply feature selection algorithms to limit the features set to a smaller size.  In the next step, the module deploys classification or estimation models to classify the data or estimate the eating durations/mass/chewing or swallowing counts. Popular classification methods in these systems include SVM, RF, and CNN whereas linear regression and logistic regression are used for estimation purpose. Furthermore, the validation component deploys different metrics to evaluate the results drawn in the previous step. Metrics like Accuracy, Precision, Recall and F1 score are utilized to evaluate classification algorithms. Authors have utilized error metrics including Absolute Mean Error, Percentage Mean Error, and Mean Square Standard Error along with some other error metrics for evaluation of estimation tasks.

Next sub section describes the above concepts in details for better understanding. 

\subsection{Eating terminologies in food monitoring}
\begin{figure*}[ht]
 \centering
 \includegraphics[scale=0.80]{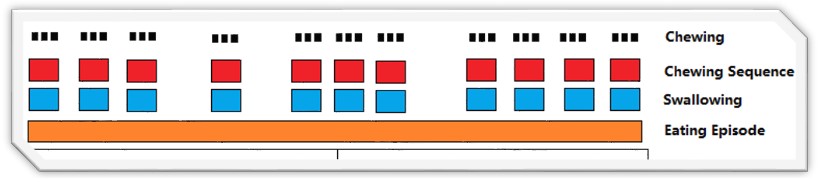}
 \caption{Chewing, Chewing Sequence, Swallowing and Eating Episode defined over a timeline}
 \label{fig:eating_timeiline}
 \end{figure*}

In this study we define \textbf{\textit{Eating}} as the process of food intake comprised of micro activities including picking, taking the food to mouth, chewing and swallowing. \textbf{\textit{Chewing}} is the process of biting the food into smaller parts in the mouth with the help of teeth, which usually creates motions in jawbone. In some studies, mastication has been used as an alternative word for chewing. A \textbf{\textit{chewing sequence}} is combination of multiple chews before swallowing takes place. \textbf{\textit{Swallowing}} is defined as the process of passing of food from mouth to stomach through throat, as an immediate step after chewing in most of the cases. An \textbf{\textit{eating episode}} is defined as the continuous chewing process followed by swallowing with small silent windows of time, whereas a long delay (more than 10 minutes for instance) defines another eating episode. Fig. \ref{fig:eating_timeiline} explains the above definitions in graphic format.

In food monitoring systems, \textbf{\textit{food intake type}} is defined as the distinct food being eaten. In some studies, researchers have classified the intake type into soft and hard type, whereas some authors classified the exact food being eaten. Eating duration is time taken for an eating episode, usually a single meal time.

\subsection{Major Sensors}  
\begin{figure*}[ht]
 \centering
 \includegraphics[scale=0.85]{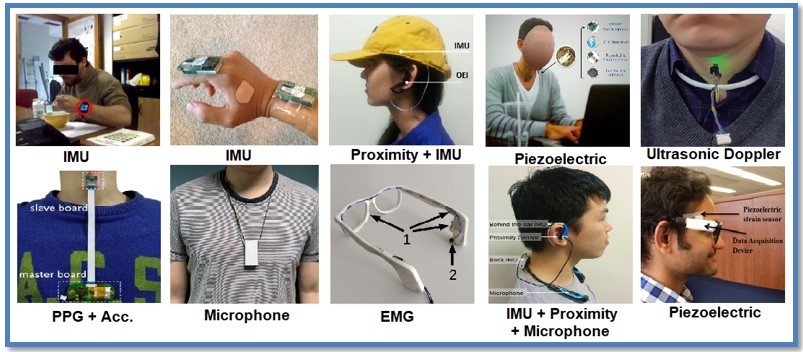}
 \caption{Major Sensors placed at different location on the body, as found in the literature}
 \label{fig:major_sensors}
 \end{figure*}

One or multiple sensors are used in automatic food monitoring systems to assist in eating activity detection, chewing/swallowing detection, food intake type detection and eating duration estimation. In some research studies, wearable devices were manufactured by utilizing sensors, processing units and power management units. However, some researchers have emphasized on using built-in sensors in smart wearable devices. Low cost has been emphasized in these studies apart from some expensive examples. Some sensors used in different approaches under the literature review conducted in this research work are given in  Fig. \ref{fig:major_sensors}. 

\textbf{\textit{IMU Sensors}} – Inertial Measurement Unit (IMU) are motion sensors which are comprised of multi-axis accelerometer and gyroscope. An accelerometer measures the acceleration of the device, whereas gyroscope measures rotational movement of the device in which these are mounted. These sensors capture the body part movement in multi-axis, and provide related features which are utilized to monitor eating activities. Researchers have used these sensors for automatic food monitoring by deploying majorly on wrist and head of human body. 

\textbf{\textit{Microphone Sensors}} – These sensors are alternatively called as acoustic sensors in automatic food monitoring systems. The microphone provides the audio features of chewing and swallowing sound, which are utilized to monitoring eating activities. Some wrist watches and bands along with wearable necklaces integrate these sensors. In some research studies microphones have been mounted on small wearable boards in different shapes. These sensors are majorly used in ear-based wearable devices, necklaces or wrist-based devices.

\textbf{\textit{Piezoelectric Sensors}} – With the ability to measure pressure, strain, acceleration or force, these sensors are used for automatic food monitoring. Miscellaneous features from vibration signals are used for chewing and swallowing detection majorly. These sensors have majorly been used behind the ear lobe, head area or neck area by the researchers.

Apart from the above mentioned sensors, the uncommon sensors in automatic food monitoring include Electromyography, PPG, ambient and some motion detection sensors. 

\subsection{Major Classification Algorithms}
The classification algorithms play an imperative role in automatic food monitoring, as data from different types of signals is taken as input, which varies significantly in different ranges. Most of the data to process is found in numeric format since signals are digitized for computer-based processing. In some cases, classification is required in binary mode i.e. chewing/non-chewing, eating/non-eating etc. in most cases. In some cases, more than two classes are defined for example [eating while walking, eating while sitting, walking, talking], [sitting, left chew, right chew, talking].  Therefore, the choice of classifier becomes imperative to deal with different kind of classification. We define some of the widely utilized classification algorithms in automatic food monitoring now.  

Researchers have used \textbf{\textit{Support Vector Machine (SVM)}} for classification, which is a machine learning classifier. It models the data in space in a way that there is a clear distance/gap between different categories/class labels. Being a supervised learning algorithm, in this algorithm data is divided into training and test sets while model is applied. SVM suites well in case of high dimensional data, however its training time is higher for large datasets.

\textbf{\textit{Random Forest (RF)}}, another supervised classification algorithm, has also been used by researchers for classification task in automatic food monitoring systems, which uses multiple decision trees for prediction task. The prediction results are obtained by the average of all decision trees utilized under calculation. While RF usually yields better prediction results than other algorithms, the algorithm tends to over fit. 

\textbf{\textit{Neural Networks (NN)}} is a deep learning classification model that is comprised of multiple nodes, where each node is connected to the other with a weight value. The weight value defines the relationship between nodes, whereas relationship between different nodes can be defined via linear combinations. This technique is useful due to the high computation power, but often requires more training data for building the model. Researchers have utilized different types of NN like LSTM Network, CNN and RNN.

\subsection{Major Evaluation Measures}
\begin{figure}[ht]
 \centering
 \includegraphics[scale=0.50]{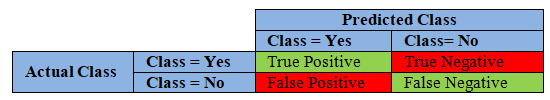}
 \caption{Confusion Matrix for Actual and Predicted Classes}
 \label{fig:confusion_matrix}
 \end{figure}

In literature, we have found that the performance of classification models is evaluated by using the following metrics; Accuracy, Precision, Recall and F-Measure (or F1 Score). These measures utilize outcomes of the classification model; True Positive (TP), False Positive (FP), True Negative (TN) and False Negative (FN). \textbf{\textit{True Positive}} is the number of cases where classifier predicted actual positive class. \textbf{\textit{False Positive}} defines the number of cases where classifier incorrectly predicted a positive class whereas the actual class was negative.  \textbf{\textit{True Negative}} is the number of cases where classifier predicted the negative class, whereas the actual class was also negative. \textbf{\textit{False Negative}} is the number of cases where the classifier predicted positive whereas the actual class was negative.  Fig. \ref{fig:confusion_matrix} defines these classes using a confusion matrix.

As described earlier, Accuracy, Precision, Recall and F-Measures are calculated using the above mentioned outcomes from the prediction model.

\textbf{\textit{Accuracy}} can be defined as the percentage of the ratio of predicted positives and all cases. Mathematically:

\begin{center}
$Accuracy = \frac{TP+TN}{TP+FP+TN+FN}*100$
\end{center} 

However, since accuracy only takes the predicted positives into account, it is generally not clear, if any false positives are impacting the accuracy score. Therefore, we find usage of Precision and Recall in literature by several researchers. \textbf{\textit{Precision}} is defined as the percentage of the ratio of True Positive cases with the all positive predicted cases. Mathematically:

\begin{center}
$Precision = \frac{TP}{TP+FP}*100$
\end{center} 

Furthermore, \textbf{\textit{Recall}} is defined as the percentage of the ratio of True positives cases with actual positive cases. Mathematically:

\begin{center}
$Recall = \frac{TP}{TP+FN}*100$
\end{center}

Another measure which uses both Precision and Recall is termed as \textbf{\textit{F-Measure}}, which becomes critical when a balance between Precision and Recall is required. It is defined as below:

\begin{center}
$F-Measure = \frac{Precision*Recall}{Precision+Recall}*2$
\end{center} 
 \begin{figure*}[ht]
 \centering
 \includegraphics[scale=0.85]{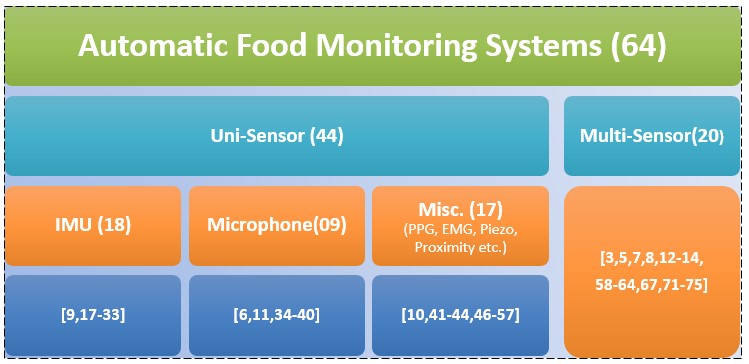}
 \caption{Sensor-based hierarchical distribution of research work done in automatic food intake monitoring systems}
 \label{fig:work_division}
 \end{figure*}
\section{Literature Review}
As depicted in Fig. \ref{fig:work_division}, automatic food activity monitoring literature can be classified by the number and types of sensors. Some approaches utilized only one sensor, whereas there has also been a trend of using more than one sensor to target better results in terms of detection accuracy and scalability for future. This section is divided into two sub sections, i.e. uni-sensor approaches and multi-sensor approaches. In the first sub-section, we divide literature by the type of sensor used by the researchers.

\subsection{Uni- Sensor Approaches}
Many studies in the past used a single sensor for automatic food monitoring. These sensors include IMU, microphone, Piezoelectric and other sensors. We provide review of major sensors in first two sections (IMU and Microphone), third section contains reviews of the approaches with miscellaneous other sensors. We also provide summary of these approaches in table, listing the main contributions, methodology, experiments and limitations, to help the critical analysis. 

\subsubsection{IMU Sensor-based Approaches}
\citeauthor{lee2019user} \cite{lee2019user} presented an approach called Instant Detection of Eating Action (IDEA) based upon a wristband sensor to monitor eating actions. The identification process works at two steps. First step, called as Generalized Model, involves application of DNN over individual user instances to classify eating actions, during which unconfirmed actions are separated. It also identifies other users having similar patterns. In the next step, which is called Personalized Model, similar eating patterns are used as training set to further classify the unconfirmed instances. Based upon the signal data, the data is segmented at the first step, which is then passed to the two tier methodology defined above for eating detection. The methodology is tested with 36 participants, with a wristband, whereas videos were recorded to obtain the ground truth. The methodology yielded 92\% F1 score, which depicts the high performance, keeping in view the usage of single sensor. However, this methodology cannot be extended towards a comprehensive solution.

To achieve eating activity detection, \citeauthor{zhang2017generalized} \cite{zhang2017generalized} carried out a methodology in wild environment using a single wrist worn inertial sensor. At the first step, the number of data points in each second window is matched against a threshold, and only reliable windows are shortlisted. The noise from data points is removed by applying a Gaussian Kernel. A two-stage approach is adapted to detect eating in the wild. In the motion detection stage, the stationary period of individuals is separated from other activities. A high-recall classifier is used to identify the stationary periods with LOPO approach. The next stage, feeding detection is adapts a motif-based approach, which has two major components. First component searches for motifs, in order to generate a candidate set, and then second component works for candidate classification. RF algorithm is used to classify the feeding and non-feeding gestures. The methodology is tested on 8 participants, whereas 1920 minutes of data was collected for experiments. The results indicate 90\% recall and 61\% F1 score. The lower F1 score with high recall indicates that the methodology is vulnerable to false positives, which requires more investigation with a large dataset. 

Another similar approach to detect eating events using IMU sensors was presented by \citeauthor{kyritsis2018end} \cite{kyritsis2018end} who utilized inertial sensors of smart watch. In this approach, signals are smoothed using 5-th order median filter and a high pass FIR filter. The next step involves classification using two networks (CNN and LSTM). The methodology is tested with a publicly available FIC dataset of 10 participants with 10 meal sessions. The approach is able to achieve 85\% Precision, 92\% Recall, and 88\% F1 score for eating event detection, in natural conditions. However, major emphasis is done on movement detection, whereas chewing and swallowing micro-activities are not targeted. 

Recently, \citeauthor{kyritsis2019modeling} \cite{kyritsis2019modeling} used inertial sensor embedded into the smart watches to detect food intake cycles. The food intake cycle is defined as the sequence of wrist micro movements that is pick food, upwards, downwards, mouth, no movement and other movement. Moreover, meal session is defined as the sequence of food intake cycles. At the first step, signals are pre-processed using 5th order median filter and high-pass FIR filter. CNN is used to estimate the micro movements in each window of sensor stream, which is further classified as food intake or otherwise, using LSTM Network. The methodology is tested with 12 subjects with 21 meal sessions. The methodology is able to identify eating cycles with 90\% precision, 93\% Recall and 91\% F1 Score. However, the methodology requires that start/end moments of meal should be known. 

Although the main target of \citeauthor{kim2016eating} \cite{kim2016eating} was to notify faster eating, but they used inertial sensors in wrist bands to classify eating activities. The axis values from accelerometer and gyroscope are received, whereas the rotation angle of 3 axis is termed as motion of bite. The methodology is tested on 15 participants which achieved 97\% accuracy and 95\% precision. However, F1 score is not reported by the authors. Moreover, the methodology only takes account of rotation angle which remains ambiguous in natural environment. 

For the purpose of chewing detection in real-time uncontrolled environment \citeauthor{stankoski2020real} \cite{stankoski2020real} used smart watches. In the first step, the stream of data collected from accelerometer and gyroscope is filtered to remove different types of noises in the data. In the next step, the data is segmented to create 15 seconds sliding windows. Different time and frequency domain features are selected from the stream of data. Mutual information of the features is calculated using a custom algorithm which is further used to select a set of features for classification process. During classification phase, non-eating data is under-sampled, to construct a data set with 65\% non-eating windows and 35\% eating windows. The classification in this approach is done using RF classifier, with LOSO Cross Validation. Based upon low precision values using RF, the predictions are smoothed using HMM.  The methodology was tested using 10 participants, 70 different meals, and 161 hours of different activities in uncontrolled real-life conditions. The approach is able to achieve 70\% Precision and 83\% Recall, which is lower than the other similar wrist-based approaches, as compared by the authors. However, authors claim that the data for other approaches was recorded within laboratory settings.  Although, the methodology is tested with uncontrolled real-life environment, but the number of false positives (low precision) need improvement.

Targeting the same goal of chewing detection, \citeauthor{wang2015care} \cite{wang2015care} proposed a system called CARE (Chewing Activity Recognition) based upon a 1-axis accelerometer. The authors claim that any accelerometer based component can be used for head/face movement tracking, for example, glasses, fillet or an earphone. The data from these sensors is passed to a mobile application for chewing identification. First step of the detection process is extraction of time domain features (8) and frequency domain features (12). These features are then passed to classification process, which implements 5 classification algorithms. The algorithms include DT, MLP, SVM, NN and WSVM. A customized formula is created to calibrate the acceleration data, which is then passed to two more algorithms called mean crossing counts (MCC), and window peak detect (WPD). These algorithms are used for chewing frequency detection, where percentage error is used for evaluation of these algorithms. The methodology is tested with Shimmer sensor, where 150 minutes' data of different activities is collected from 4 participants. According to authors, the WSVM classifier provides balanced performance with 96\% accuracy, 91\% precision, 92\% Recall and 91\% F Score. For chewing counts, the WPD algorithm worked better than MCC algorithm. However, the applicability of technique to a wide variety of participants with real-time environment conditions is lacking. 

\citeauthor{maramis2016real} \cite{maramis2016real} also worked on chewing detection using 3-axis orientation data acquired by a smart watch. In the first step, the segments are annotated with either bite or control instances. These segments are further smoothed using B-Spline Interpolation followed by re-sampling. A rolling disk background removal algorithm is further applied before the features are extracted from the signals. The extracted features are passed to the SVM classifier for classification. The methodology is applied on 8 participants, who are given milk with cereals. The process yielded 92\% accuracy for bite detection from the segments, however, the precision and recall are unknown. Moreover, only hand movements are considered as eating moments, which limits the enhancement of this technique towards chewing and swallowing detection.

An eating episode detection methodology was presented by \citeauthor{kyritsis2017automated} \cite{kyritsis2017automated} who used inertial sensors. The approach detects micro-movements and then creates eating episodes by combining sequences of micro-movements. Micro-movements include pick food, move upwards, move mouth downwards, no movement and other movement. In the first step, the data from sensor streams is smoothed using 5th order mean filter. In the next step, the gravity component is removed from accelerometer signals. Features extraction process extracts time and frequency domain features from the signal data. SVM Classifier is used to detect the micro-movements. HMM is further used for eating episode detection. The data of 8 subjects is used for testing purposes, with 109 minutes' data. The approach resulted in 78\% Precision and 77\% Recall, which is lower than other approaches using same sensor. 

The same authors modified this approach by utilizing different set of algorithms in \cite{kyritsis2017food}. Like before, authors divide the eating activity into micro-movements. After detection of micro movements, the methodology detects event intake movements in a meal session. Signal smoothness, gravity removal and feature extraction is done in similar fashion as in previous approach. The micro-movement detection is done using Support Vector Machine (SVM) classifier. Furthermore, LSTM Network is used to classify intake and non-intake cycles in the eating process. A dataset from university restaurant for 10 subjects is utilized by the authors to test the approach, with a total duration of 130 minutes.  The approach achieves 88\% precision, 91\% recall, and 89\% F1 score. The results appear promising for eating activity detection but rely heavily on movements therefore its extendibility towards chewing and swallowing detection is limited.

Eating episode detection was also targeted by \citeauthor{thomaz2015practical} \cite{thomaz2015practical} whose proposed an inertial sensor based technique to recognize eating in controlled and uncontrolled environments. In this first step, 3-axi accelerometer sensor data is received from a smart watch. The sensor data is filtered using exponentially-weighted moving average (EMA) filter, and frames are extracted from the stream. For each frame, different features are extracted, and passed to food intake classification process. Random Forest Algorithm is used for classification of food intake gesture, which are further grouped together to estimate eating moments using DBScan algorithm.  Three case studies are performed to test the approach, i.e 20 participants with semi-controlled lab conditions, 7 participants in wild conditions (one day) and 1 participant in wild-long conditions (31 days). The first wild day case study with 7 participants yielded 76\% F1 score, whereas the 31-days wild day case study with 1 participant yielded 71\% F1 Score. The precision in both cases is around 65\% which indicates that 35\% of the predictions are false positives, higher than other studies in this domain. 

In another similar approach \citeauthor{kyritsis2019detecting} \cite{kyritsis2019detecting} used smart watch sensors to detect eating moments in free living conditions. In the first step, the signals from accelerometer and gyroscope are smoothed using a moving average filter which is further passed through a Finite Impulse Response Filter. The eating detection is done using end to end NN. The methodology is tested with 28 participants out of which 16 participants were tested in free-living conditions. The approach provides 90\% precision, 89\% Recall and 89\% F1 Score, which is slightly lower than other similar approaches with same sensor.  

In a different methodology \citeauthor{rahman2015unintrusive} \cite{rahman2015unintrusive} proposed to use head motion to detect eating actions. For this purpose, Google Glass has been utilized to capture the motion with built-in accelerometer and gyroscope. In the first step data from sensors is received, and several features are extracted and passed to classification module which applies KNN, NB, C45 and RF classifiers. This methodology has been tested with 38 participants, who performed a series of activities like eat, drink, walk and talk etc. in controlled environment. The results suggest that the methodology is able to achieve 67\% F0.5 score with Random Forest Classifier. A close inspection into the precision results of RF Classifier reveal that the methodology was failed to detect eating for 24\% of the participants, whereas the precision is between 40\%-80\% for 40\% of the participants. These results show that the system behaves in unstable manner in controlled environment. Moreover, the high cost glass puts a negative impact on its usage.

Another similar approach which uses eye glasses was presented by \citeauthor{farooq2018accelerometer} \cite{farooq2018accelerometer} who attached an accelerometer to the frame of regular eyeglasses for eating detection. At the first stage the high-filter is used on the signals to remove DC component and normalize the signal. An easting sequence is obtained with different second-frames (3 to 30 secs). Against each eating sequence, a set of 38 features is extracted from the 3-axes signal data. These features are passed to Minimum Redundancy and Maximum Relevance (mRMR) and Forward Feature Selection (FFS) to selected most relevant features. KNN classifier is used in the next step with 10-fold cross validation to classify the sequences in eating or non-eating classes. Ten participants are used to test this methodology in free-living and controlled conditions. The combined data resulted in better performance for 20sec eating window, where F1 score was measured up to 88\%. Authors claim that these results are similar to some other approaches with multiple sensors with un-natural way of wearing. However apart from the compulsion of intrusive eyeglasses, the approach treated liquid as food intake along with solid intake, where the percentage of liquid is unknown. It’s also tested on 10 participants only with unknown number of hours, therefore more experimentation is required. 

Conversely, \citeauthor{lin2019comparison} \cite{lin2019comparison} argue that the movement axes of wrist, hand and finger are congruent; therefore the fingers can be as good as wrist for detection of food. In this research work a comparison of wrist and finger-based inertial sensors is drawn for chewing detection. Different features from the inertial sensor are collected and sent to a customized algorithm, for both cases for bites detection. This methodology is tested with 10 meals against a single person with 170 minutes' data. The results, in terms of detection accuracy, TP and FP show that both location (wrist, finger) yield same results. However, this methodology is tested with only 1 person, and thus can't be generalized based upon these results. 

Based on the same perception, \citeauthor{fan2016eating} \cite{fan2016eating} deployed a 3-axis accelerometer and gyroscope mounted ring on fingers to capture the inertial data for the purpose of eating detection. The received data is segmented at the first step using motion energy calculator. 22 features of the data are passed to classification process to categorize the windows as eating or non-eating using algorithms including SVM, NB, KNN, RF, DT, NN and Log. R. The methodology is tested by deploying a sensor mounted ring and wrist sensor whereas the number of subjects is unknown; however, the reported gestures count to 375. KNN performs better than other algorithms with 91\% accuracy. Although the accuracy of KNN using ring-based sensor than using KNN for wrist-based sensor, however critical measures like precision and recall are not reported. 

In a unique study, \citeauthor{san2020eating} \cite{san2020eating} used a small inertial sensor mounted wireless device attached to the underside of jawbone for eating detection. 4-second windows are regarded as chewing, whereas 20-second windows are regarded as chewing bouts, and a group of chewing bouts within 2 minutes is termed as an eating episode. In the first step, the data is normalized with z-score measure, and 4-second windows are defined. The next step involves feature extraction against each frame, and RF Classifier is used for eating detection. Furthermore, DBScan is used to group the chewing data into chewing bouts, and adjacent chewing bouts within 120 seconds duration are termed as eating episodes. The methodology is tested with 14 participants. Within lab conditions, the precision for three types of detections is above 80\%, however, the recall is 47\% for chewing detection in laboratory conditions. The eating episode detection in wild environment has 92\% precision and 89\% recall. The substantial differences between different measures in different conditions suggest that the approach needs to be tested with a larger population. Moreover, the awkward position of sensor is un-natural in general. 

Another unique study was presented by \citeauthor{wang2018eating} \cite{wang2018eating} who mounted an accelerometer in a headband for muscle contraction sensing for the purpose of eating detection. The work includes eating activity detection and chews count detection. Noise from signals is removed using median filter at the first step, whereas segmentation involves a sliding window operation to create window segments. Afterwards 23 features are extracted from the signal which is further normalized using z-score algorithm. The next step applies 5 classification algorithms namely DT, NN, MLP, SVM and WSVM for eating activity detection from other activities speaking, sitting, standing, walking, drinking and coughing. The methodology is applied to 4 subjects in controlled conditions. It is revealed that the average accuracy of the algorithms is 94\% and F1 score is 87\% for chewing detection. The chewing counting algorithm provides 12\% average error. The results are encouraging however due to small dataset size, the approach requires more testing in natural environment, along with the intrusive nature of the headband.

A summary of these approaches is given in TABLE \ref{tab:summary_imu},  which details the main contribution, the activities monitored and limitations.
\begin{table*}[t]
\caption{Summary of IMU-based approaches for food intake monitoring systems}
\label{tab:summary_imu}
\centering 
\begin{tabular}{p{0.5cm}p{8cm}p{1.0cm}p{6cm}}
\hline\hline 
\textbf{Ref.} & \textbf{Main Contributions} & \textbf{Activities} & \textbf{Limitations} \\ [0.5ex] 
\hline 
\cite{stankoski2020real} & Real-time eating detection using a Smart watch inertial sensor &	CD & Low Precision\\
\cite{lee2019user} &Eating activity detection using wrist band time series data.	&EAD&	Limited to eating activity detection only.\\
\cite{zhang2017generalized} &Eating detection using wrist band IMU sensor data&	EAD&Low F1 Score with high Recall\\
\cite{kyritsis2018end}&Eating micro-movements detection using inertial sensor in smart watches	&EAD	&Low F1 Score than other approaches with IMU Sensor\\
\cite{kyritsis2019modeling}&Food intake cycle detection using inertial sensor data&	EAD&	Manual Input of start and end of meal\\
\cite{kim2016eating}&Eating moment detection to indicate faster eating using inertial sensor&EAD&	Unknown Recall and F1 Score. \\
\cite{wang2015care}&Chewing detection using single axis accelerometer	&CD&	Analysis with variety of participants/ data is required.\\
\cite{maramis2016real}&Bite instance detection using smart watch orientation data&	CD	&Unknown Precision, Recall. \\
\cite{kyritsis2017automated}&Eating episodes detection using wrist band IMU sensor data&EED&	Precision and Recall is lower than other approaches. \\
\cite{kyritsis2017food}&Food intake detection from inertial sensors	&EED&	Relies heavily on movements so extendibility towards chewing/swallowing is limited.\\
\cite{thomaz2015practical}&Eating moments recognition using wrist based inertial sensor data&	EED&	Low Precision value\\
\cite{kyritsis2019detecting}&Eating moments detection using inertial data from smart watches&	EED&	Low F1  Score than other approaches with IMU sensors\\
\cite{rahman2015unintrusive}&Eating activity detection using Google Glass.	&EAD	&Low Accuracy and Precision.\\
\cite{farooq2018accelerometer}&Eating activity detection using accelerometer in free living conditions&	EED&	Intrusive\\
\cite{lin2019comparison}&Bite detection using finger motion and comparison with wrist based detection&	CD	&Tested with 1 participant only.\\
\cite{fan2016eating}&Eating gestures detection using finger motion detection	&EAD&	Unknown Precision, Recall, and dataset Size.\\
\cite{san2020eating}&Eating episode detection using smart watch inertial sensor&	EED	&Intrusive approach, Testing in larger environment required\\
\cite{wang2018eating}&Chewing detection using mastication muscle contraction sensor&	CD&	Intrusive, slightly lower F1 Score than other approaches.\\
\hline
\end{tabular}
\end{table*}

\subsubsection{Microphone-based Approaches}
Researchers in the past have developed uni-sensor approaches using microphone sensor as well. In these approaches, microphone is utilized in different wearable devices for different activities detection in automatic food monitoring paradigm. For instance, \citeauthor{gao2016ihear} \cite{gao2016ihear} used commercially available Bluetooth headset to utilize the microphone sensor for chewing detection, whereas the methodology is termed as iHearFood. Authors argue that chewing the harder foods generate more sound as compared with soft foods, therefore these sound patterns are helpful in determining the food type. The received audio is processed to created shorter segments, and time and frequency domain features are extracted from the signals. The selected features only include the simple features which do not involve high computation. In the next step, SVM and RMBs classifiers are used for classification of data. The methodology is tested with 28 participants, where they were asked to perform activities like eating, speaking and drinking. The SVM classifier was able to achieve 95\% accuracy in laboratory settings, and up to 76\% accuracy in wild settings under different conditions. The RBM classifier achieved accuracy up to 94\% in the wild environment. However, the precision and recall values are not reported, which are critical to observe the underline true/false predictions.

Another ear-worn microphone based approach was presented by \citeauthor{papapanagiotou2017chewing} \cite{papapanagiotou2017chewing} for chewing detection. At the first step, a low-pass filter is applied on audio signals for down sampling to lower frequency, followed by an FIR high-pass filter to remove very low frequencies. The window frames with 1, 2, 3 and 5 seconds are gathered for classification purpose. CNN is applied on the data to classify the instances as chewing or non-chewing. This methodology is tested with 14 participants in semi-free living conditions with 60 hours' data. It appears that the methodology provides more than 95\% accuracy for 2-5 second windows, whereas the F1 score is calculated to be 88\%. The results suggest that the approach can further be tested in full living conditions.

Recently, \citeauthor{bi2018auracle} \cite{bi2018auracle} designed a microphone based ear device to detect chewing sound. At the first step the features from time and frequency domain are collected from microphone signal. Down sampling is used to balance the data, and 700 features from 62 categories are extracted. Recursive Free Elimination Algorithm is used to select 40 features for the classification process. 3-second windows are classified as chewing/non chewing using Logistic Regression, whereas JSC and Wards Metrics are used for Eating Episodes detection.  The methodology is tested with 14 participants in free living conditions. IT was found that Logistic Regression provides 93\% accuracy, 76\% precision, 81\% Recall and 78\% F1 score. The eating episodes detection accuracy was found to be 77\% and 93\% using JSC and Ward's Metric respectively. Importantly the Episode based results include 12 False Detections for both algorithms, showing weakness of the approach. Moreover, F1 score is lower than some other approaches using microphone sensor. 

Conversely, \citeauthor{thomaz2015inferring} \cite{thomaz2015inferring} argue that the ambient sounds from Smartphone microphones or wrist based microphones can be targeted to detect eating activities, whereas the most difficult task in such process is to identify information-rich features from the audio data. Authors process the audio signals at the first step to create audio frames of size 50ms using Hanning filtered sliding window. A set of 50 features is extracted from each frame using Yaafe tool, which mainly include time and frequency domain features like Zero Crossing rate, loudness, energy, spectral variation etc. Furthermore 400 frames are merged together to creation a 10 second frame. In the next step, RF classifier is used for eating/non-eating classification. The methodology is tested with 21 participants, which resulted in 80\% F1 Score. The results achieved by this methodology are lower than some other approaches, along with the issue of miss classification for shorter meals and audio privacy concern which is un-addressed. 

Using same sensors, \citeauthor{kalantarian2015audio} \cite{kalantarian2015audio} presented an approach to classify eating activities (bites) from other activities. At the first step of the process, the signals are received, and pre-processed for feature extraction. An open source speech and music interpretation tool named openSMILE is used for feature extraction purposes. Audio based features like Signal Energy, Loudness, Pitch, Line Spectral Patterns, Zero Crossings, Duration and Peaks are extracted along with other low-level audio features. A total of 6555 features are passed to Feature Selection phase. The feature selection phase utilizes Correlation Feature Selection Subset Evaluator technique to choose best features suitable for classification. RF Classifier is used to classify the bites in the data. The methodology is tested with 10 subjects, who are asked to perform different activities include eating (chips, apples), Noise, Water intake and Talk. The methodology yielded 87\% precision and recall for eating activities detection. It is not evident if the approach is able to detect eating episodes, apart from bites detection. Moreover, only two types of food are tested, in controlled conditions. Therefore, more generalization is required along with scalability testing for other eating related activities like episodes detection, swallowing and eating durations. 

On the other hand, \citeauthor{shin2019accurate} \cite{shin2019accurate} argue that usage comfort and accurate eating detection in wild environments are two most important characteristics of automated diet monitoring systems. The authors propose a wearable necklace mounted with a microphone sensor to detect eating episodes, with firm contact with the skin of the user. At the first step, raw signal data is standardized, 3 second windows are created and 23 features (out of 900 features) for each window are selected. The windows are classified as eating and non-eating using DNN classifier. The methodology is tested with 2 participants by collecting 90 minutes' data.  Authors show the accuracy up to 86\%, whereas precision and recall are not known. Apart from the inadequate testing and limitation of sensor's firm attention to skin, the detection results are comparatively lower than other approaches.

In a similar approach, \citeauthor{turan2018detection} \cite{turan2018detection} used a throat microphone to detect chewing and swallowing. Three types of events are defined at the start namely chew, swallow and rest. At the first step down sampling is performed followed by a high-pass FIR filter to remove the noise from the signals. CNN is further used for classification of the frames. The methodology is tested with 8 participants in laboratory settings. The methodology is able to achieve 78\% accuracy and 78\% F1 Score, which can be termed as in-adequate based upon other similar approaches. 

Another neck-based approach was presented by \citeauthor{bi2015autodietary} \cite{bi2015autodietary} in which a microphone is used for capturing eating sounds, and is wore around the neck area.  At the first step, the amplification and filtration is performed to ensure the better quality of signals whereas these are digitized in the next step. Afterwards frames are created through the segmentation process. HMM is used to detect chewing and swallowing. For food type classification, 34 features from time domain and frequency domain are extracted from these frames, whereas the classification is done using DT classifier. The methodology is tested with 12 subjects. The chewing/swallowing accuracy is calculated as 87\%, whereas food type classification accuracy yielded 87\% accuracy, 88\% Recall, 86\% Precision and 87\% F1 Score. Apart from this, 100\% (rounded) F1 score is reported for liquid and solid food classification.   However, the experiments are conducted only in laboratory conditions.

In a unique research work \citeauthor{kondo2019robust} \cite{kondo2019robust} proposed a method to classify eating data, collected in the form of sounds received from bone conduction microphone. The proposed process extracts 75 features from the received data at the first step. 5-fold cross validation is applied using different classification algorithms including DT, SVM, NN and Ensemble Classifiers. The methodology was tested on 9 participants by collecting 95 minutes' data. The data was comprised of 2001 chewing samples, 504 talking samples, 151 swallowing samples and 118 misc. samples. The obtained results revealed that SVM provides better accuracy (97\%) as compared with the other algorithms. The algorithm accuracy decreases to 92\% with 7 features, whereas by using 14 features the accuracy drops to 95\%. The classification accuracy of meal-related activities yields good results in terms of accuracy, however precision and recall results are not available to further evaluate the approach keeping in view that SVM is generally sensitive to class imbalance. 

A summary of these approaches is given in TABLE \ref{tab:summary_microphone} which details the main contribution, the sensors used in the approach, the activities monitored and limitations.

\begin{table*}[t]
\caption{Summary of Microphone-based approaches for food intake monitoring systems} 
\label{tab:summary_microphone}
\centering 
\begin{tabular}{p{0.5cm}p{7cm}p{2cm}p{5cm}}
\hline\hline 
\textbf{Ref.} & \textbf{Main Contributions} & \textbf{Activities} & \textbf{Limitations} \\ [0.5ex] 
\hline 
\cite{bi2018auracle} & Chewing and Eating Episode detection using microphone based Ear device	& CD, EED &	Low F1 score for chewing detection, Low Accuracy with False detections for Eating episodes\\
\cite{gao2016ihear} & Eating activity detection using Bluetooth microphones	& EAD	& Unknown Precision and Recall.\\
\cite{papapanagiotou2017chewing} & Chewing detection using microphone and CNN	& CD	& Wild environment testing missing\\
\cite{thomaz2015inferring} & Eating activity detection using ambient sounds in wild environment	& EAD	& Low F1 Score, audio privacy concern un-addressed\\
\cite{kalantarian2015audio} & Eating activity detection using smartwatch microphone &	EAD	& Limited food types are tested during experiments\\
\cite{shin2019accurate} & Eating episodes detection using wearable necklace	& EAD	& Low Accuracy, Unknown Precision and Recall.\\
\cite{turan2018detection} & Chewing and Swallowing detection using throat microphone	& CD, SD	& Low Accuracy and F1 Score\\
\cite{bi2015autodietary} & Chewing, swallowing, and food type detection using microphone	& EAD, CD, FTD	& No experimentation is done in Wild\\
\cite{kondo2019robust} & Eating micro-activities detection using microphone in real environment	& CD, SD	& Unknown Precision and Recall\\
\hline 
\end{tabular}
\end{table*}

\subsubsection{Miscellaneous Approaches}
Although IMU and microphone sensors are widely used in uni-sensor approaches, however a relatively small number of approaches used sensors like EMG, Piezoelectric strain, and PPG sensors etc.

\citeauthor{zhang2020retrieval} \cite{zhang2020retrieval} argue that the timing errors during eating detection are better metrics of evaluation that the other metrics like Accuracy and F1 Score. Moreover, the authors consider a bottom up approach rather than top bottom approach for eating detection. In bottom up approach, the chewing cycles are detected first, and drilled further to detect chewing segments contrary to top-bottom approaches. In this study, the authors used diet glasses with Electromyography(EMG) sensor targeting temporalis muscles sensing.  The approach uses the proposed bottom-up algorithm works in 6 steps. In the signal pre-processing phase, a notch filter is applied in order to eliminate the power line interference on the EMG data. A high-pass filter is applied to remove the motion arti-crafts and baseline wander. In the next step, chewing cycles are detected using the EMG onset detection principle as proposed by 20. Furthermore, the chewing segments are detected using a custom algorithm applied on the chewing cycles. Two more algorithms named as Fusion of multi-source detection and Gap Elimination is applied on chewing segments for fine tuning purposes. The algorithm is tested with two top bottom algorithms namely Threshold-based algorithm and ocSVM detection algorithm. The approach is tested with 122 hours' data of 10 young participants in free living conditions, along with the other two algorithms. The proposed algorithm achieved 99\% F1 Score, better than other two algorithms. The algorithm also outperformed the other two algorithms in terms of start/end time errors, which are measured as 2.4 +- 0.4s and 4.3 +-0.4s. This research has better performance than both of the other two similar works. The approach requires testing with older population for generalization purpose, as timing errors are linked to eating motions in different ages.  

Another approach based on Electromyography (EMG) was presented by \citeauthor{zhang2016diet} \cite{zhang2016diet} for chewing monitoring, whereas the EMG electrodes are attached to the frame of eyeglasses. The authors particularly emphasized upon detection of chewing cycles and classification of food texture. The recorded chewing cycles are pre-processed using a band-pass filter to remove fluctuation and noise. Furthermore, the methodology adapts two algorithms for chewing detection process. For food classification, RF and LDA are used with 10-fold cross validation.  The methodology is tested on 8 participants, whereas different food was given to the participants to record the data of 5345 chewing cycles. The authors reported 80\% precision and recall for chewing detection using signal-energy based detection algorithm. The food classification mean accuracy is 57\%, whereas the precision and Recall are unknown. Although the idea of using EMG electrodes in eyeglasses is good, but better classification algorithms for chewing and food classification based upon EMG data should be experimented along with adapting a feature selection mechanism. 

\citeauthor{zhang2018free} \cite{zhang2018free} also used EMG-based eyeglasses, which record temporalis muscle activity, to separate eating and non-eating moments in free-living conditions by. Moreover, the authors estimate the start and end time of eating episodes. At the first step, 6 features from EMG time series are selected, which are normalized using L2-Normalizer. Afterwards ocSVM is used for classification purpose. The methodology was tested on 10 participants in free-living conditions, which results in 95\% F1 score for eating/non-eating classification. The approach yields better results than other approaches with different other sensors, but requires intrusive eyeglasses during eating. 

In another similar approach, \citeauthor{zhang2017monitoring} \cite{zhang2017monitoring} used smart eyeglasses with Electromyography (EMG) electrodes for chewing cycle detection. At the first step EMG data is filtered using a notch filter in order to remove power line interference followed by a high pass filter to remove motion artefacts. Chewing detection is performed by a method adapted by \cite{der1998detection}.  The methodology is tested with 10 participants in laboratory conditions (118 minutes' data) and free living conditions (429 minutes' data).  Chewing detection resulted in 94\% precision and 94\% recall in lab conditions which declined to 79\% and 77\% in free living conditions. The laboratory results are encouraging; however, the free living conditions yield low precision and recall. 

Instead of EMG sensor, \citeauthor{farooq2016automatic} \cite{farooq2016automatic} used piezoelectric film sensor behind the earlobe for detection and quantification of chewing. The signals from sensor are demeaned first, and then processed through a low-pass filter to remove noise. Signals are smoothed using a moving average filter. ANN is applied on fixed size window signals to classify the eating episodes and non-eating episodes. A chewing algorithm is applied on eating episodes to quantify the chewing counts.  The algorithm mainly utilized number of peaks within the signals to estimate the chew counts. To detect peaks within signals, a threshold-based peak detection algorithm is used whereas threshold is calculated using histograms considering signal amplitudes. The counts were compared with the manual chew counts and errors in both approaches were found to be same.  The methodology is tested on 30 participants, whereas 110 distinct food items were used in the experiments. The experiments recorded 60 hours data, whereas 26 hours' data was related to food intake. The data included 5467 chewing episodes (counted manually) with 62001 chews. The approach resulted in 91\% F1-Score for ANN Classification, whereas the mean absolute error for chewing count algorithm was found to be 15\%. The mean absolute error is compared with other approaches, showing that the methodology results in almost similar error rates.  The interesting results suggest that the methodology can be tested in wild environment along with a more natural way to wear the sensor. 

In another study, \citeauthor{farooq2016linear} \cite{farooq2016linear} worked out on a linear regression model in order to estimate chews from chewing segments, whereas chewing segments are identified using piezoelectric strain sensor placed near ear. In the first stage, the authors applied low-pass filter to remove high frequency components from the signals, whereas 4 features are computed against each chewing sequence, using the peaks in the signals. Five different models are developed for estimation purpose, where 4 models were created with individual features and one model was developed using all features. The methodology was tested on 30 subjects, and different meals were assigned to the subjects including breakfast, lunch and dinner. Moreover, each participant was asked to perform meal activity with a 5-minute rest before and after the activity. 5467 chewing sequences were obtained with 62001 chews. It was found that the individual feature models resulted in error rate between 10.15\%-13.96\%, whereas the model with all features resulted in 9.66\% error. The estimation results are encouraging however the chewing episode detection remains manual in the process, which can be automated. Moreover, the model can be improved by adding more features from the sensor data. 

\citeauthor{alshurafa2015recognition} \cite{alshurafa2015recognition} designed a necklace mounted with piezoelectric sensor along with a Smartphone app for swallow and food type detection. The onboard microcontroller in the necklace samples the voltage of the sensor, converts the voltage to digital signal and transmits the data to the mobile app. At the second step, a swallow detection algorithm is utilized. A sliding window is applied for waveform generation, which are smoothed via Saitzky-Folay Convolution filter. The number of swallows is detected by counting the number of peaks, using a threshold for spacing between the swallows. A spectrogram is generated in the next step using Short-time Fourier Transform. The spectrogram is used to extract 360 features, which are reduced to 30 using a feature selection algorithm. The next phase involves classification via KNN, Bayesian Network and Random Forest. The methodology is tested on 10 subjects in the first experiment. The solid food detection resulted in 94\% recall and 94\% precision using Random Forest classifier (better than other classifiers). Furthermore, the food type detection (between Sandwich and chips) resulted in 72\% recall and 72\% precision. Experiment 2 included 20 participants, whereas the performance decreased in terms of precision and recall (both averaged at 87\%) for solid and liquid detection using Random Forest Classifier. However, hard food classification with nuts, chocolate and patty resulted in 80\% precision and recall which is higher than experiment 1. The varying results suggest that more experiments are required for generalization purpose for both swallow detection as well as food type detection along with ability to test in natural environment. Moreover, the methodology requires proper attachment of necklace with the skin for good results, apart from the fact that it will intrusive for people to wear a necklace.

Instead of necklace, \citeauthor{farooq2016segmentation} \cite{farooq2016segmentation} proposed to use piezoelectric sensor on temple of glasses to detect eating from temporalis muscle movement. At the first step signals are sampled using a 12-bit ADC followed by amplification using ultra low power amplifier. Capacitors enable noise removal and high voltage elimination. Afterwards segmentation is done by Hanning Sliding window to identify high energy signal segments. SVM is used for chewing classification using three features. For chewing count estimation Linear Regression is used. The methodology is experimented with 10 participants in Laboratory and Free living conditions, and results were collected by merging both datasets. The experiments yielded 95\% precision, 98\% recall and 96\% F1 score. The chewing count estimation algorithm yielded 3.83\% Average Mean Absolute Error (AMAE). The above average F1 score for chewing detection suggest the efficiency of the approach compared with other sensors. However, like other eye-glasses approaches, there is a difficulty in adaptation.

In a unique approach \citeauthor{wang2020wieat} \cite{wang2020wieat} used channel state information (CSI) extracted from smartphones or WiFi-enabled IoT devices for eating activity detection, along with chewing and swallowing estimation. In the first step, the CSI measurements are received from the WiFi device of the user. The received data is pre-processed to remove different types of noise. Afterwards, the segmentation process based upon spectrograms determines the activities start and end time. From different activities, eating activity is separated using K-Means clusters. Two more modules are designed to detect the utensils, and chewing/swallowing estimation. In the first module, a series of steps is adapted to identify the utensils being used by the user. These steps include feature extraction, classification using SVM, and probability based technique to identify the utensil. The second module counts the chews and swallows with a series of steps. These steps include reconstruction of CSI motion within the window, chewing period estimation using frequency domain, and threshold-based swallow detection. The methodology is tested with 20 participants, and 1600 minutes eating data. Authors are able to achieve eating activity classification accuracy up to 92\%, eating motion with utensils accuracy up to 93\%, and 13\% percentage error in chewing and swallow estimation. However, in real-world scenario, the technique remains very burdensome to adapt as normally user’s eating place changes in the course of time. Moreover, if the user is surrounded by other people during eating, the detection process fails, as it is common problem in WiFi signal-based people identification. It becomes even more complex and error-prone, if more than one user is using this technique in an eating place, for example a dinner/lunch party, or dinner/lunch with family members. 

In a recent study, \citeauthor{schiboni2018privacy} \cite{schiboni2018privacy} targeted the privacy invasive approach for camera based eating detection. The authors fixed a camera on a cap, which pointed downwards during eating. The field of view for camera is configured using a formula which ensures that the dietary objects are filmed only. The methodology additionally discards images raising privacy concerns, while ensuring the required food images are captured at the same time. By using the video recordings, for dietary event identification, the DNN algorithm is deployed. 3 second windows are spotted as eating/non-eating based upon the presence of dietary object in all frames in the window. The authors reported 90\% Recall for eating event detection; however, the methodology is tested with 1 participant only. The methodology, on one hand relies heavily on dietary objects which exist in different shapes, colors and sizes, whereas on the other hand cannot be generalized based upon results from 1 participant. Apart from this, the approach requires people to wear a hat/cap for images capturing purpose. 

In order to detect chewing, \citeauthor{papapanagiotou2016novelconf} \cite{papapanagiotou2016novelconf} proposed to monitor jaw and muscle motion by using photoplethysmography (PPG) sensor mounted on an earphone placed at human body's outer ear. The sensor data is pre-processed using a high-pass FIR filter and further normalized based upon the control signals received from the amplifier. The chewing detection is performed by three algorithms; Maximum Sound/Signal Energy (MSEA), Low-Pass Filtering Algorithm (LPFA) and Chewing-Band Power Algorithm (CBPA). The methodology is tested with 21 participants who performed different activities like talking, coughing, eating etc. The results from three algorithms suggest that MSEA performs best for snacks detection in terms of precision (93\%) and Recall (92\%), however the chews detection precision is as low as 71\% and Recall is as low as 40\%. Other algorithms also result in unstable precision and recall for Chews, Bouts and Snacks detection, therefore, algorithms like SVM or RF can be experimented, as these widely used by other researchers.

In another study \citeauthor{farooq2015comparative} \cite{farooq2015comparative} aimed to compare the performance of piezoelectric strain sensor and plotter drawn strain sensor (lower in cost) for chewing count detection.  The author targeted chewing counts estimation within eating episodes, which were marked manually. At the first step, the signals were demeaned for inter-subject variability. Furthermore, the signals were smoothed using a low pass filter. Afterwards a chewing count estimation algorithm was applied using data from both sensors, and mean absolute error was calculated to compare the performance of sensors. The methodology is tested with 5 participants with different activities including rest, talking, reading, eating and walking. During the experiments the sensors were placed below left and right ears. 98 chewing episodes were recorded with 2488 chews. After the chewing counts were obtained along with absolute mean errors in estimation, a statistical test was applied to compare the performance. It was concluded that there was no difference in performance, whereas error rate remained around 8\% for both sensors. The approach provides interesting results, and further investigation using printed strain sensor can be executed as only 5 participants were used in experiment. 

In a unique approach \citeauthor{hotta2017eating} \cite{hotta2017eating}, measured heart rate data using a smart watch to utilize for eating detection. In this first step, the data is processed to remove noise, and smoothed using median filter. The next step involves segmentation using sliding window, and features are extracted for classification. The classification in this approach is done using SVM classifier for eating moment detection. The approach is tested with 9 participants, which resulted in 99\% accuracy, 95\% precision, 41\% recall and 57\% F1 Score. The low recall and F1 scores indicate that the performance of the approach is lower than other approaches. 

\citeauthor{steimer2016portable} \cite{steimer2016portable} used an air pressure sensor mounted into an ear bud targeting changes in the in-ear air pressure to detect chewing and swallowing. Different time and frequency domain features are used for classification, which is done by a binary classifier. This methodology is tested with 3 participants, however contrary to other approaches; evaluation is done using specificity and sensitivity. Specificity stands at 65\% and 28\% for swallowing and chewing respectively. The sensitivity yielded 64\% and 88\% score for swallowing and chewing respectively. The low scores for these measures show the in-adequate performance of this approach. 

Recently, \citeauthor{chun2018detecting} \cite{chun2018detecting} used a proximity sensor embedded in a necklace to classify eating and non-eating activities. At the first step chewing is detected, whereas continuous chewing frames are grouped together to formulate chewing bouts, and the last step combines multiple bouts to define eating episodes. The data received from the sensor is pre-processed for smoothing, and 5-second windows are defined. This data is passed to the next module which deploys level crossing algorithm based upon amplitude and number of crossings over a threshold to categorize the windows as eating and non-eating. DBScan is used to group the windows to define chewing bouts. A pause duration threshold is used in next step to define eating episodes. The methodology is tested over 32 participants in controlled laboratory/field conditions and wild conditions. The approach yielded 95\% precision and 82\% Recall in controlled conditions, which dropped to 78\% and 73\% respectively in wild conditions. The designed necklace position is required to point towards jawbone for correction detection, as well as the system confuses walking with chewing leading to dropped precision and recall in wild conditions.

In a different approach, the \citeauthor{chung2017glasses} \cite{chung2017glasses} used load cells on the glasses frame to monitor temporalis muscles activity. An amplified force is obtained on the hinge by using the lever mechanism between the head piece and the temple. Different features from the temporal and spectral domains are extracted out of which 84 features are used for classification with SVM classifier. The methodology is tested with 10 participants. The chewing detection yielded 90\% F1 score with 89\% precision and 90\% recall. The approach achieves 90\% F1 score in real life conditions, however it is intrusive in nature.

\citeauthor{lee2017food} \cite{lee2017food} argue that the Doppler frequency shifts can be spotted in a signal when an ultrasonic wave is formed with fixed frequency near jaw area of the person eating something. These shifts are due to masticatory movements involving chewing and swallowing. Authors developed device with two ultrasonic transmitters emitting an ultrasonic tone with two sensors to receive these signals. Chewing in this methodology is detected by monitoring jaw movements and wallowing by using hyoid bone movement. These sensors are supposed to be present under the chin and the throat front. A multi-channel digital audio inference is used to digitize the signal which is processed for feature extraction to extract four features per frame. The classification is done using ANN algorithm which takes multiple frames as input. The methodology is tested on 10 subjects. The methodology yielded 91\% and 78\% accuracy for chewing and swallowing detection respectively. The maximum F1 score for chewing and swallowing was calculated to be 91\% and 75\% respectively. The food classification yielded 91\% F1 score for chewing based classification and 77\% for swallowing based classification. However, the continuous emitted signal’s effect on human body requires further investigations. 

A summary of these approaches is given in TABLE \ref{tab:summary_misc} which details the main contribution, the sensors used in the approach, the activities monitored and limitations.

\begin{table*}[t]
\caption{Summary of Misc. Uni-Sensor based approaches for food intake monitoring systems}
\label{tab:summary_misc}
\centering 
\begin{tabular}{p{0.5cm}p{6cm}p{3cm}p{1cm}p{5cm}}
\hline\hline 
\textbf{Ref.} & \textbf{Main Contributions} & \textbf{Sensors} & \textbf{Activities} & \textbf{Limitations} \\ [0.5ex]
\hline
\cite{farooq2016automatic}  & Chewing count and rate detection	 & Piezoelectric	 & EAD, CD & 	Wild Testing Required\\
\cite{zhang2020retrieval} &  Time Performance of Chewing based eating event detection	 & EMG & 	EAD	 & Age specific dataset, Intrusive in nature\\
\cite{zhang2016diet}  & Chewing monitoring using Eye Glasses & 	EMG & 	CD	 & Low Precision and Recall\\
\cite{zhang2018free}  & Eating event spotting using EMG-mounted &  eye glasses	 & EMG	EAD & 	Intrusive\\
\cite{zhang2017monitoring}  & Chewing cycle detection using EMG based eye glasses & 	EMG	 & CD & 	Low Precision and Recall in free living conditions\\
\cite{farooq2016linear}  & Chewing count estimation & Piezoelectric &  CD  & 	Manual Chewing Episode detection\\
\cite{alshurafa2015recognition}  & Food Type detection using piezoelectric sensor embedded in a wearable necklace 	 & Piezoelectric	 & FTD & Unstable results in different conditions\\
\cite{farooq2016segmentation}  & Chewing detection and count estimation using piezoelectric sensor mounted on eye-glasses	 & Piezoelectric & CD & Wild Testing Required\\
\cite{wang2020wieat} & Eating detection, chewing and swallowing estimation using Wifi Signals	 & Wifi 	 & EAD, CE, SE & 	Hard implementation, Multi-user identification issue.\\
\cite{schiboni2018privacy} &  Eating event monitoring in free living using wearable video camera	 & Video Camera	 & EAD	 & Intrusive, Tested with only 1 participant\\
\cite{papapanagiotou2016novelconf}  & Chewing detection using PPG sensor	 & PPG & 	CD, EAD, EED	 & Low Precision and Recall\\
\cite{farooq2015comparative}  & Chewing count estimation comparison between piezoelectric strain sensor and plotter drawn strain sensor	 & Plotter Drawn Strain Sensor	 & CD	 & Small size dataset\\
\cite{hotta2017eating}  & Eating Moment detection using heart rate fetched from smart watches	 & Heart Rate data from Smartwatch & 	EAD	 & Low Recall and F1 Score\\
\cite{steimer2016portable}  & Eating speed estimation using air pressure sensor & 	Air pressure sensor & 	CD, SD	 & Un comparable results with other approaches due to different evaluation measures used.\\
\cite{chun2018detecting}  & Eating detection using jawbone movements tracking in wild environment	 & Proximity	 & EED & Intrusive Approach. Low Precision and Recall in Wild Conditions.\\
\cite{chung2017glasses}  & Chewing detection using ball type load cell mounted on eye glasses & 	Ball type load cells & 	CD & 	Obtrusive in nature\\
\cite{lee2017food}  & Chewing, Swallowing/Food type detection using Ultrasonic Doppler Sonar	 & Ultrasonic Doppler Sonar & 	FTD, CD, SD	 & Low swallow detection F1 Score \\
\hline 
\end{tabular} 
\end{table*}
 
\subsection{Multi-Sensor Approaches}
\citeauthor{bedri2020fitbyte} \cite{bedri2020fitbyte} targeted a generalized approach (FitByte) for eating detection with high performance in real time conditions. The main contributions of this research work include a single sensor-embedded eye glasses with ability to capture food intake actions. It also includes a data processing engine capable to detect eating moments, and record food images which further help to identify the food type using an algorithm.  FitByte was designed to detect different motions including jaw motion, chewing and swallowing. It can also capture food images to assist the food type identification. Proximity Sensor is used to identify hand-to-mouth gestures. Gyroscope and accelerometer sensors are used to detect chewing and swallowing, whereas a mini spy camera is used to capture food images. Data of these sensors is processed through segmentation, feature extraction and activity detection modules. RF classifier is used to classify the activities in the categories including eating, drinking, walking, talking, and silence (or no activity) using LOSO cross validation.  The system was tested with 23 participants, and provides eating detection accuracy up to 94\%, 89\% F1 Score and duration estimation accuracy up to 96.3\%. The approach has many sensors including an accelerometer, 5 gyroscopes, 1 proximity sensor, and a camera, which are installed on an obtrusive eye glasses. Removing some of the sensors, results in low performance. This Approach requires people to keep wearing special and costly eye glasses. The glasses are uncomfortable due to heavy size and unnatural for most of them. 

Targeting a single device, \citeauthor{zhang2020necksense} \cite{zhang2020necksense} designed a multi-sensor necklace mounted with a proximity sensor, ambient light sensor and IMU sensor. Using these sensors, four types of signals are received including proximity, energy, LFA and ambient light signals. Two algorithms are used in Segmentation process, namely Prominence-based peak finding algorithm, and longest period subsequence algorithm, to find chewing sequences and sub sequences respectively. In the next step, 257 features from every sequence are extracted. Authors used Friedman's GBM classifier for detection of chewing sub sequences. DBSCAN is further applied on sub sequences to cluster them in chewing episodes. The methodology is tested with 20 participants using 10 participants for exploratory study (271 hours' data) and 10 for free-living study (193 hours' data). The free-living study achieved 74\% F1 score for chewing detection and 77\% F1 score for episode level detection. The performance is low as compared to other approaches keeping in view the usage of multiple sensors, and an un-natural necklace as a requirement for detection.

\citeauthor{min2018audio} \cite{min2018audio} argue that the ear-worn devices are good options for dietary monitoring due the placement on human body. The authors explored the usage of inertial and microphone based ear-bud prototype for eating detection. In the first step, the time-domain and frequency-domain features are expected from inertial sensors, whereas MFCC features are extracted from the microphone. These features are used with three classifiers namely RF, SVM and NB Classifiers. The methodology is tested with 8 participants, which shows that RF performs better than other classifiers in terms of accuracy (73\%) for chewing detection. Not only the results are un-satisfactory (low accuracy with unknown precision and recall), the methodology needs extensive testing.

Conversely,  \citeauthor{anderez2018hierarchical} \cite{anderez2018hierarchical} suggest that the bi-nodal inertial units can efficiently be used for detection of eating and drinking. For this purpose, they deployed a tri-axial accelerometer on waist to detect physical states (standing, sitting, moving), whereas another tri-axial Micro Electro Mechanical System is put on wrist to detect eating and drinking gestures. For state detection, 6 features from un-filtered stream of signals are used with RF classifier for classification purposes. Furthermore, for gesture recognition, LSTM classifier is utilized keeping in view the sequential nature of the gestures. State detection yielded 100\% accuracy, whereas the gesture recognition achieved 99\% classification performance. However, the dataset size has not been mentioned by the authors, and the results are termed as preliminary, which require extensive testing.

In another study, \citeauthor{bedri2017earbit} \cite{bedri2017earbit} used proximity, inertial and microphone sensors for chewing and swallowing detection. First two sensors are placed in ear area for chewing detection, whereas the third sensor is placed around the neck to detect swallowing. Another IMU sensor was used behind the user's neck to detect large body motions. At the first step data from gyroscope is smoothed via a low pass filter and segmentation is done with 30 second sliding window. A forward floating selection algorithm (SFFS) is used to filter the features and 34 features are selected for classification task. Chewing is detected using Random Forest Classifier whereas continuous chewing intervals are passed through a custom process based upon some rules to mark these episodes. The methodology is tested with 16 participants in 5 sessions, where each session lasted for 75 minutes. The results show 91\% and 80\% F1 score for chewing detection in semi-controlled laboratory and free living conditions respectively. The eating episode detection accuracy was found to be 100\% and 94\% for semi controlled laboratory and free living conditions respectively. The eating episode detection can be automated, apart from the fact that the F1 scores for the same are not provided. Moreover, the approach is intrusive due to multiple sensors.

\citeauthor{papapanagiotou2016novel} modified their previous approach to include a microphone sensor with the PPG sensor to detect eating events termed as chewing bouts \cite{papapanagiotou2016novel}. Both sensors are embedded into a single wearable ear-phone. The technique also includes an accelerometer to assist the chewing detection process. Different features from microphone sensor, accelerometer sensor and PPG sensor are extracted. The SVM classifier is applied on the selected features for microphone sensor and PPG sensors separately. The results from these two sensors are combined with some features of accelerometer sensor to classify chewing bouts. The methodology is tested with 22 subjects, with 60 hours' data, out of which eating corresponds to 7.6 hours. The authors show that combining the two sensors achieves accuracy up to 94\%, which also includes additional accelerometer sensor for detecting chewing bouts (not individual chews). Moreover, the F1 score is also low (76\%) compared with other approaches, considering usage of 3 different sensors. 

Authors in this approach \cite{zhang2016bite} attached a bone vibration sensor to their EMG-mounted eyeglasses proposed before, to target chewing detection and food classification. No classification results are presented, however authors claim that the EMG waveforms and bone vibrations indicate that the data can be used for chewing detection and food classification respectively. Moreover, this approach is not tested with laboratory or wild condition environment. 

In order to detect eating in wild environment, \citeauthor{farooq2016novel} \cite{farooq2016novel} proposed piezoelectric strain sensor and accelerometer sensor. The piezoelectric strain sensor is placed at temporalis muscle and accelerometer is placed on temple of eyeglasses for detection process. A feature set is obtained from these sensors at first step to help the classification process, with the assumption that eating episode energy is higher than non-eating episodes. Moreover, accelerometer signals have higher energy in physical activities as compared to sedentary activities. Activity classification is done using two approaches. The first approach combines the features from both sensors to create a single vector, which is passed to multiclass linear SVM. The other approach works in two step. First step uses two classifiers where the first classifier (an SVM) detects food intake using piezoelectric sensor signals and second classifier (an SVM) uses accelerometer signals to differentiate between walking and static categories. A decision tree is implemented in next step to estimate the final class of the activity. The methodology is tested on 10 participants, whereas a data of total 2185 episodes is collected with different activities (322 eating episodes). The First approach results in 95.7\% Average F1 Score, whereas the second approach results in 99.85\% Average F1 Score. The stable prediction scores indicate the methodology needs experimentations in un-restricted free living conditions,  however it requires eyeglasses uses along with additional sensor attachment to temporalis overhead. 

In another study, \citeauthor{farooq2016detection} \cite{farooq2016detection} used 3 sensors for food detection including a piezoelectric film sensor (attached to ear), a hand-to-mouth gesture sensor (wrist-based), and a tri-axial accelerometer (neck based). In the first step, the signals are processed using high-pass filtering and normalized afterwards. All signals are divided into 10 second frames, whereas 68 features including time and frequency features are selected. All features are scaled before the classification phase. Authors applied DT, LDA, and Logistic Regression for classification purposes.  The methodology is tested on 12 subjects, in free living conditions. Authors reported highest accuracy achieved using LDA Classifier with 93\% score, whereas precision and recall values stand at 97\% and 90\% in this case. However, the 3 sensors make it difficult in terms of acceptance by general public, along with the unknown robustness towards more specific eating detection activities like chewing, swallowing, and food type detection.

\citeauthor{bedri2015detecting} \cite{bedri2015detecting} presented a methodology which uses outer ear interface to distinguish between eating and non-eating activities. The methodology requires 3 infrared proximity sensors embedded in the ear bud and a gyroscope sensor placed in the hat. Five features are selected from the sensor data (2 from proximity sensors, 3 from gyroscope) which are passed to HMM for classification purpose. The 5 second frames were classified as Eating or Non Eating as a result with accuracy, precision and recall values. The methodology is tested in laboratory conditions with 23 participants. The results suggest that the precision and recall scores improve as the window size gets larger, whereas the precision and recall stands at 93\% for 5-minute windows.  The wild testing is done using 6 participants, whereas decline is monitored for precision which falls to 42\%. The detailed investigation by authors led to conclusion that the classifier mixed talking with eating in most cases, resulting in lower precision. Although the results are encouraging, however authors suggest adding a microphone to control precision values which could add negatively to an already intrusive approach. 

\citeauthor{bedri2015wearable} \cite{bedri2015wearable} also argue that proximity sensors are low powered, vigorous to noise, and privacy preserving, thus can effectively be used for eating detection. Authors in this research work designed a system called Outer Ear Interface (OEI). The authors embedded 3 proximity sensors in an earphone, targeting the monitoring of jaw movements during eating. The setup also includes an IMU inside a hat for body movement tracking. The activities are detected by training a set of HMM using the features extracted from sensors. The methodology is tested with 23 participants in laboratory conditions to detect four types of activities: eating, silent, talking and walking. The system yielded 95\% accuracy, 93\% precision and 96\% recall for eating activity detection. Apart from wearing a cap to accommodate the IMU, the experiments are done in laboratory settings, and thus the applicability to real-world environment is not known. Moreover, the system’s scalability towards chewing and swallowing detection is unknown. 

Human activity recognition \cite{yao2018human,hong2019variant} are often employed for food intake monitoring. \citeauthor{hossain2019human} \cite{hossain2019human} explored the use of eSense sensors and supervised machine learning algorithms for activity recognition \cite{hong2018disturbance}, majorly divided into head and mouth-related activities. The device used in the research contains a microphone and 3-axis accelerometer. The data from the device is received, and time domain features are extracted, which are further used for classification using ensemble learning algorithms \cite{chen2010multiobjective,chen2009regularized}, i.e. random forests, SVM, KNN and CNN. To test the methodology, a dataset of 50467 records is obtained. The activities include speaking, eating, handshaking, head nodding, staying and walking. The CNN algorithm achieved 100\% accuracy for eating activity; however, the precision and recall values are not available to further measure the performance of the approach. Moreover, it is not known whether the approach was experimented in natural environment or laboratory conditions. 

In order to find the reasons of difference in eating detection accuracy between laboratory and free living conditions, \citeauthor{mirtchouk2017recognizing} \cite{mirtchouk2017recognizing} adapted a multi-sensor approach consisting of audio and motion sensors. For laboratory data collection, authors used Google Glass, Earbud, Smart watch (both wrists), whereas for free-living conditions, the Google Glass was not used.  In the first step, signals from different sensors are pre-processed and features are extracted. The methodology extracts 14 features from audio signals and 32 features from motion sensors. Random Forest classifier is used to classify the meal frames. In lab conditions 59 hours' data was collected with 12 participants, whereas in free-living conditions, 5 participants were involved and the data of 112.5 hours was collected. The accuracy achieved in both experiments exceeds 90\%, however, the highest precision in all cases is 45\%.

\citeauthor{mirtchouk2016automated} \cite{mirtchouk2016automated} also argue that audio and motion features best suited to food type detection due to their complementary nature. More specifically, authors hypothesize that motion sensors are good for soft food detection, whereas audio sensors can better be used in food texture recognition. Four sensors are used in this methodology including microphone, a smart watch, Google glass and 3 IP based video cameras. In the first step, noise is removed from the signals and audio and motion features are selected from the sensors. The food type classification is done using RF classifier, whereas linear regression is used for weight estimation. The methodology is tested with 6 participants in laboratory conditions.  It has been found that audio provides an accuracy up to 67\%, whereas the head/wrist motions provide 76\% accuracy. Using all sensors, authors are able to achieve 83\% classification accuracy for food type detection, whereas the absolute relative error was measured to be 35\% during weight estimation. It provides enough evidence that multiple sensors result in better performance. However, the food type classification accuracy cannot be trusted without knowledge of precision and recall.
 
 \begin{table*}[t]
\caption{Summary of Multi-Sensor based approaches for food intake monitoring systems}
\label{tab:summary_multi}
\centering 
\begin{tabular}{p{0.5cm}p{6cm}p{4cm}p{1.5cm}p{4cm}}  
\hline\hline
\textbf{Ref.} & \textbf{Main Contributions} & \textbf{Sensors} & \textbf{Activities} & \textbf{Limitations} \\ [0.5ex]
\hline 
\cite{bedri2017earbit} & Chewing and Swallowing detection using proximity, inertial and microphone sensors	& Proximity, IMU, Microphone	& CD, SD	& Intrusive\\
\cite{doulah2017meal} & Comparison of performance of AIM , Food Diary and Push Button approaches for eating duration estimation	& Piezoelectric, Proximity, Accelerometer	& EDE	& Obtrusive\\

\cite{farooq2016detection} & Food detection using piezo electric, gesture recognition and tri-axial accelerometer sensors		& Piezoelectric, hand-to-mouth gesture sensor, accelerometer	& CD &	Intrusive\\
\cite{mirtchouk2017recognizing} & Multi-sensor approach to spot eating detection in laboratory and free living conditions	& Audio and motion sensors, Google Glass	& EAD	& Low Precision\\
\cite{zhang2016bite} & Usage of bone vibration sensor and EMG mounted eye glasses for chewing detection	Bone Vibration Sensor, & EMG Sensor	& CD	& No Evaluation results\\
\cite{farooq2016novel} & Usage of Piezoelectric and accelerometer for eating detection in Wild environment		& Piezoelectric, accelerometer		& EAD		& Experimentation is required in Free Living Conditions, Intrusive Approach\\
\cite{yang2019statistical} & Mass and Energy estimation using video and chewing sensor	& piezoelectric sensor, Video Camera	& EE	& Sensor attachment to skin is required. \\

\cite{bedri2020fitbyte} & Eating episode detection and duration estimation using multiple sensors 	& IMU, Proximity, Mini Spy Camera	& EED, EDE	& Intrusive Approach with many sensors, Removal of any sensor results in lower performance\\
\cite{zhang2020necksense} & Multi-sensor necklace for chewing and eating episode detection in Free Living Conditions	& IMU, Proximity, Ambient Light	& CD, EED	& Intrusive Approach , Low F1 Score\\
\cite{min2018audio} & Inertial and Microphone mounted ear bud for eating detection	& IMU, Microphone	& CD	& Low Accuracym, Unknown Precision, Recall\\
\cite{anderez2018hierarchical} & Eating and drinking recognition using accelerometer and tri-axial micro electro mechanical system	& Accelerometer Tri-axial micro electro mechanical system	& EAD	& Unknown dataset size\\

\cite{papapanagiotou2016novel} & Chewing detection using PPG, Audio and Accelerometer 	& PPG, Microphone, Accelerometer	& CD	& Low F1 Score\\

\cite{bedri2015detecting} & Mastication detection using proximity and gyroscope sensors		& Proximity, Gyroscope		& EAD	& 	Low precision in free living conditions\\

\cite{bedri2015wearable} & Proximity and IMU-based eating activity recognition	& Proximity, IMU  & 	EAD	 & Intrusive  \\ 
\cite{hossain2019human} & Eating activity detection using microphone and accelerometer embedded into an earbud	& Microphone, Accelerometer	& EAD	& Unknown Precision and Recall, Unknown testing environment\\

\cite{mirtchouk2016automated} & Food type detection using body-worn audio and motion sensors	& Microphone, IMU, Google Glass, Video  & Camera	FTD	& Unknown Precision and Recall\\

\cite{farooq2019validation} & Comparison of Food Intake detection performance of multiple sensors approach and video system	& Piezoelectric, Proximity, Accelerometer	& CD, EAD	& Low F1 score despite multiple sensors\\

\cite{zhang2018habits} & Eating detection using wearable necklace	& Ambient Light Sensor, Proximity Sensor, IMU	& EED, CD	& No Experiments conducted\\
\cite{farooq2017real} & Eating activity classification using piezoelectric and accelerometer mounted on eye-glasses	& Piezoelectric, Accelerometer	& EAD	& Obtrusive\\
\cite{li2018novel} & Swallow defection using throat acceleration sensor and PPG Signals	& PPG, Accelerometer	& SD	& Only tested with drinks, Intrusive\\
\hline
\end{tabular}
\end{table*}

In a recent study, \citeauthor{yang2019statistical} \cite{yang2019statistical} targeted mass and energy intake estimation at meal level by utilizing a piezoelectric strain sensor attached below the outer ear with capability to capture jawbone movements along with a video camera. At the first step 57 features are extracted from the sensors. Using these features two statistical regression models are developed for meal and energy intake estimation. Using forward selection method, based upon mean absolute percentage error, 6 features were selected for the regression model. The methodology is tested with 30 participants in laboratory conditions which resulted in 250.8 Root Mean Square Error. It will be interesting to evaluate this technique in wild environment.   

\citeauthor{farooq2019validation} \cite{farooq2019validation} conducted a study which aimed comparing the ability of sensors used \cite{farooq2016detection} with a multi camera video system. Authors targeted the validation of activity detection and food intake detection. The feature extraction and classification is done using the same approach as in \cite{farooq2016detection}. Whereas Light's Kappa statistic is measured for Video Annotation in order to draw a comparison. The methodology is tested on 40 participants in unconstrained environment. The methodology yielded 80\% F1 Score for activity recognition, and 80\% F1 Score for chewing detection, compared with Annotation Kappa values 77\% and 76\% respectively. 1-Way ANOVA suggests that there is no difference between both approaches. 

\citeauthor{doulah2017meal} \cite{doulah2017meal} tested their previous system \cite{farooq2016detection} for three tasks including duration of eating episode, actual digestion duration and number of eating events.  Same set of sensors, as in \cite{farooq2016detection} is used in this approach. Two more systems are utilized for comparison purposes namely food diary and push button \cite{sazonov2008non}.  The methodology has been tested with 12 participants, with 23 eating episodes. It was found that for duration of eating episode, there was difference between proposed system and food diary and push button, whereas proposed system and push button had similar results. 

\citeauthor{zhang2018habits} \cite{zhang2018habits} designed a necklace which imbeds three sensors including an ambient light sensor, a proximity sensor and an IMU sensor. Signals from all sensors are saved into memory for processing, and spectrograms are generated. The spectrograms show that the approach can be used to characterize eating activities; however, the authors have not represented any experiments. 

Recently, \citeauthor{li2018novel} \cite{li2018novel} presented a swallowing detection technique using throat-based accelerometer and Photoplethysmogram (PPG) signals. Both signals are processed using an algorithm. The PPG algorithm extracts two features (entropy and maximal signal energy) from the 1-second windows of PPG Signal, whereas the ACC Algorithm extracts time domain features from accelerometer signal windows. SVM is applied on both signals to detect swallowing, and results from both signals are used in a Logistic Regression model to estimate the swallowing. The LR model is based upon the training data. The methodology is applied on 20 participants, resulting in 91\% precision and 60\% specificity. The comparative results suggest that the performance is better when detection is combined from both sensors. However, this methodology is difficult to adapt in natural conditions as well as it has only been tested with drinks. 

In another multi-sensor based study, \citeauthor{farooq2017real} \cite{farooq2017real} used a piezoelectric and accelerometer to detection eating activities in real time whereas both sensors are placed on eyeglasses. At the first step, the signals from sensors are divided into 5 seconds windows and 5 features are computed. These features are passed to DT classifier to classify into 4 categories namely “eating while sitting”, “sedentary”, “eating while walking” and “walking” categories. The methodology was tested with 10 participants. The average F1 score for both eating categories was found to be 93\% with 93\% average precision and 94\% average recall. It will be interested to extend that approach towards chewing detection and swallowing detection keeping in view the usage of piezoelectric sensor. However, the eye-glasses are intrusive and inapplicable in some cases where people use mandatory eye-glasses, unless modified specially.

A summary of these approaches is given in TABLE \ref{tab:summary_multi}which details the main contribution, the sensors used in the approach, the activities monitored and limitations.

\begin{table}[t]
\caption{Division of literature with respect to activities monitored}
\label{tab:lit_division}
\centering
\begin{tabular}{p{3cm}p{5cm}}
\hline\hline
\textbf{Activity} & \textbf{Papers} \\ [0.5ex] 
\hline 
Eating Activity Detection & \cite{mirtchouk2017recognizing,gao2016ihear,farooq2016novel,lee2019user,zhang2017generalized,kyritsis2018end,kyritsis2019modeling,kim2016eating,rahman2015unintrusive,fan2016eating,thomaz2015inferring}
\cite{kalantarian2015audio,shin2019accurate,zhang2018free,schiboni2018privacy,papapanagiotou2016novel,hotta2017eating,bedri2020fitbyte,anderez2018hierarchical,bedri2015detecting,bedri2015wearable,hossain2019human,farooq2019validation}\\
Chewing Detection & \cite{farooq2016detection,stankoski2020real,farooq2016automatic,wang2015care,maramis2016real,lin2019comparison,san2020eating,wang2018eating,papapanagiotou2017chewing,turan2018detection,zhang2016diet}
\cite{zhang2017monitoring,farooq2016linear,alshurafa2015recognition,papapanagiotou2016novel,zhang2020necksense,min2018audio,papapanagiotou2016novel,farooq2019validation,li2018novel}\\
Swallowing Detection & \cite{steimer2016portable,li2018novel}\\
Eating Episode Detection & \cite{kyritsis2017automated,kyritsis2017food,thomaz2015practical,kyritsis2019detecting,farooq2018accelerometer,san2020eating,chun2018detecting,zhang2020necksense}\\
Mass/Energy Estimation & \cite{yang2019statistical,mirtchouk2016automated}\\
Food Type Identification & \cite{mirtchouk2016automated,zhang2016diet,alshurafa2015recognition}\\

\hline
\end{tabular}
\end{table}

\section{Discussion on Literature Review}
In this section we present discussion on the techniques reviewed in the literature, as detailed in section III. We classify these techniques on the basis of monitored activities, sensors used and location of the sensors for better understanding. We also provided detailed analysis of each type of approaches using different parameters including sensor location, testing environment, accuracy, precision, recall and F1 score. At the end of this section we provide future guidelines based upon our analysis of these approaches. 

It is evident from TABLE \ref{tab:lit_division} that most approaches targeted chewing detection, as it is one of the basic steps for food monitoring systems. Moreover, swallowing has also been targeted by some of the approaches. We can see that Eating duration, food mass estimation and food type detection are core areas in which the future work can be done as there has not been much work done in these areas as per our review results. 

Now we provide the summary of experiments held by all of the approaches with similar categorization of techniques as in the literature review section above. We present results of these approaches by dividing these into targeted detection categories. More specifically we divide these into Eating Activity Detection, Eating Episode Detection, Chewing Detection, Swallowing Detection, Food Type Detection and Estimation. 

We look into IMU-based approaches first as detailed in TABLE.  \ref{tab:comparison_imu}. Eating activity detection has been reported to achieve 92\% F1 Score by utilizing the IMU sensor on wrist with DNN Classifier, however the testing environment is not known in this case. It can also be seen that Eating Episode detection has yielded up to 92\% F1 Score by using IMU sensor placed on eyes, with KNN Classifier tested in Laboratory conditions. Another approach for living conditions achieves upto 91\% F1 Score. Moreover, chewing detection has achieved 91\% F1 score by using IMU sensor placed at head, with WSVM classifier, however testing environment is unknown in this case.

\begin{table*}[t]
\caption{Comparison of different evaluation measures for IMU-based approaches}
\label{tab:comparison_imu}
\centering
\begin{threeparttable}
\begin{tabular}{p{0.5cm}p{2cm}p{4cm}p{1.5cm}p{1cm}p{1cm}p{1cm}p{1.5cm}} 
\hline\hline 
\textbf{Ref.}	& \textbf{Location}	& \textbf{Algorithm}	& \textbf{Environment}	& \textbf{Accuracy}	& \textbf{Precision}	& \textbf{Recall}	& \textbf{F1 Score} \\ [0.5ex] 
\hline 

 \multicolumn{4}{l}{\textbf{Eating Activity Detection}}\\
 \hline
\cite{lee2019user} & Wrist	& DNN	& SFL	& NA	& NA	& NA	& \textbf{92\%}\\
\cite{zhang2017generalized} & Wrist	& RF	& FL	& NA	& NA	& 90\%	& 61\%\\
\cite{kyritsis2018end} & Wrist	& LSTM	& L	& NA	& 85\%	& 92\%	& 80\%\\
\cite{kyritsis2019modeling} & Wrist	& LSTM	& FL	& NA	& 90\%	& 93\%	& 91\%\\
\cite{kim2016eating} & Wrist	& No Algo.	& L	& 97\%	& 95\%	& NA	& NA\\

\cite{rahman2015unintrusive} & Eyes	& RF	& L	& NA	& NA	& NA	& 67\%(0.5)\\
\cite{fan2016eating} & Hand	& KNN	& NA	& 91\%	& NA	& NA	& NA\\
\hline
\multicolumn{4}{l}{\textbf{Eating Episode Detection}}\\
\hline
\cite{kyritsis2017automated} & Wrist	& HMM	& FL	& NA	& 78\%	& 77\%	& 78\%\\
\cite{kyritsis2017food} & Wrist	& LSTM	& FL	& NA	& 88\%	& 91\%	& 89\%\\
\cite{thomaz2015practical} & Wrist	& RF,DBSCAN	& SFL	& NA	& 67\%	& 89\% &	76\%\\
\cite{kyritsis2019detecting} & Wrist	& NN	& FL	& NA	& 90\%	& 89\%	& 89\%\\

\cite{farooq2018accelerometer} & Eyes	& KNN	& L	& NA	& 90\%	& 94\%	& \textbf{92\%}\\
   &        & 		& FL	& NA	& 87\%	& 77\%	& 86\%\\
   \cite{san2020eating} & Jawbone	& RF	& FL	& NA	& 92\%	& 89\%	& 91\%*\\

\hline
\multicolumn{4}{l}{\textbf{Chewing Detection}}\\
\hline
\cite{stankoski2020real} & Wrist	& HMM	& FL	& NA	& 70\%	& 83\%	& 76\%*\\
\cite{wang2015care} & Head	& WSVM	& NA	& 96\%	& 91\%	& 92\%	& \textbf{91\%}\\
\cite{maramis2016real} & Wrist	& SVM	& L	& 92\%	& NA	& NA	& NA\\
\cite{lin2019comparison} & Finger	& Custom	& NA	& 79\%	& 96\%	& NA	& NA\\
\cite{san2020eating} & Jawbone	& RF	& L	& NA	& 80\%+	& 47\%	& NA\\
\cite{wang2018eating} & Head	& DT, NN, MLP, SVM, WSVM	& L	& 94\%	& NA	& NA	& 87\%\\

\hline 
\end{tabular}
\begin{tablenotes}\footnotesize
\item[*]F1 score is calculated by current authors based on precision and recall values.
\end{tablenotes}
\end{threeparttable}
\end{table*}

\begin{table*}[t]
\caption{Comparison of different evaluation measures for Microphone-based approaches} 
\label{tab:comparison_micro}
\centering 
\begin{threeparttable}
\begin{tabular}{p{0.5cm}p{2cm}p{2cm}p{1.5cm}p{1cm}p{1cm}p{1cm}p{1.5cm}} 
\hline\hline 
\textbf{Ref.}	& \textbf{Location}	& \textbf{Algorithm}	& \textbf{Environment}	& \textbf{Accuracy}	& \textbf{Precision}	& \textbf{Recall}	& \textbf{F1 Score} \\ [0.5ex] 
\hline 

\multicolumn{4}{l}{\textbf{Eating Activity Detection}}\\
 \hline
\cite{gao2016ihear} & Ear	& SVM	& L	& 95\%	& NA	& NA	& NA\\
 &		& SVM	& FL	& 76\%	& NA	& NA	& NA\\
 & 		& RBM	& FL	& 94\%	& NA 	& NA	& NA\\
 \cite{thomaz2015inferring} & Wrist	& RF	& FL	& NA	& 90\%	& 76\%	& 80\%\\
\cite{kalantarian2015audio} & Wrist	& RF	& L	& NA	& 87\%	& 87\%	& \textbf{87\%}\\
\cite{shin2019accurate} & Neck	& DNN	& FL	& 86\%	& NA	& NA	& NA\\
\cite{bi2015autodietary} & Neck	& HMM	& L	& 87\%	& NA	& NA	& NA\\

\hline
\multicolumn{4}{l}{\textbf{Food Type Detection}}\\
\hline

\cite{bi2015autodietary} & Neck(Diff. Foods)	& DT	& L	& 85\%	& 84\%	& 86\%	& 85\%*\\
	& Neck (Liquid)	& DT	& L	& 100\%	& 100\%	& 100\%	& \textbf{100\%}\\
	& Neck (Solid)	& DT	& L	& 100\%	& 100\%	& 100\%	& \textbf{100\%}\\
\hline
\multicolumn{4}{l}{\textbf{Chewing/Swallow Detection}}\\
\hline

\cite{bi2018auracle} & Ear	& LR	& FL	& 93\%	& 76\%	& 81\%	& 78\%\\
\cite{papapanagiotou2017chewing} & Ear	& CNN	& SFL	& 95\%	& NA	& NA	& \textbf{88\%}\\
\cite{kondo2019robust} & Head	& SVM	& FL	& 97\%	& NA	& NA	& NA\\

\hline
\multicolumn{4}{l}{\textbf{Eating Episode Detection}}\\
\hline
\cite{bi2018auracle} & Ear		& JSC	& FL	& 77\%	& NA	& NA	& NA\\
	&	& Ward's	& FL	& 93\%	& NA	& NA	& NA`\\
\cite{turan2018detection} & Throat	& CNN	& L	& 78\%	& NA	& NA 	& \textbf{78\%}\\

\hline 
\end{tabular}
\begin{tablenotes}\footnotesize
\item[*]F1 score is calculated by current authors based on precision and recall values.
\end{tablenotes}
\end{threeparttable}
\end{table*}

\begin{table*}[t]
\caption{Comparison of different evaluation measures for Misc. Uni-Sensor based approaches} 
\label{tab:comparison_misc}
\centering 
\begin{threeparttable}
\begin{tabular}{p{0.5cm}p{3cm}p{2cm}p{2cm}p{1.5cm}p{1cm}p{1cm}p{1cm}p{1.5cm}} 
\hline\hline
\textbf{Ref.}	& \textbf{Sensor} & \textbf{Location}	& \textbf{Algorithm}	& \textbf{Environment}	& \textbf{Accuracy}	& \textbf{Precision}	& \textbf{Recall}	& \textbf{F1 Score} \\ [0.5ex] 
\hline 

\multicolumn{4}{l}{\textbf{Eating Activity Detection}} \\
\hline
\cite{zhang2020retrieval} & EMG	& Eyes	& Custom Algo	& FL	& NA	& NA	& NA	& \textbf{99\%}\\
\cite{zhang2018free} & EMG	& Eyes	& OCSVM	& FL	& NA	& NA	& NA	& 95\%\\
\cite{wang2020wieat} & Wifi	& -	& SVM	& L	& 92\%	& NA	& NA	& NA\\

\cite{schiboni2018privacy} & Camera	& Head	& DNN	& NA	& NA	& NA	& 90\%	& NA\\
\cite{papapanagiotou2016novelconf} & PPG	& Ear	& MSEA	& NA	& NA	& 93\%	& 92\%	& 93\%\\
& &			& LPFA	& NA	& NA	& 64\%	& 99\%	& 78\%\\
& &			& CPBA	& NA	& NA	& 82\%	& 97\%	& 89\%\\
\cite{hotta2017eating} & NA	& Wrist	& SVM	& NA	& 99\%	& 95\%	& 41\%	& 57\%\\
\hline
\multicolumn{4}{l}{\textbf{Eating Episode Detection}}\\
\hline

\cite{papapanagiotou2016novelconf} & PPG		& Ear	& MSEA	& NA	& NA	& 93\%	& 92\%	& \textbf{93\%}\\
\cite{chun2018detecting} & Proximity	& Neck	& LCA	& L	& NA	& 95\%	& 82\%	& 88\%*\\
	& & & 			& FL	& NA	& 78\%	& 73\%	& 75\%*\\

\hline
\multicolumn{4}{l}{\textbf{Chewing Detection}}\\
\hline
\cite{farooq2016automatic}  & Piezoelectric	& Behind Ear Lob	& ANN	& L	& NA	& NA	& NA	& 91\%\\
\cite{zhang2016diet} & EMG	& Eyes	& SEBA	& NA	& NA	& 80\%	& 80\%	& 80\%\\
\cite{zhang2017monitoring} & EMG				& Eyes	& Custom Algorithm	& L	& NA	& 94\%	& 94\%	& 94\%\\
     &					&	&			& FL	& NA	& 79\%	& 77\%	& 78\%*\\
\cite{farooq2016segmentation} & Piezoelectric	& Eyes	& SVM	& SFL	& NA	& 95\%	& 98\%	& \textbf{96\%}\\

\cite{papapanagiotou2016novelconf} & PPG	& Ear	& MSEA	& NA	& NA	& 71\%	& 40\%	& 51\%\\
		& & 	& LPFA	& NA	& NA	& 53\%	& 76\%	& 62\%\\
		& &	& CPBA	& NA	& NA	& 63\%	& 57\%	& 60\%\\
\cite{chun2018detecting} & Ball Type Load Cells		& Eyes	& SVM	& NA	& 89\%	& 94\%	& 94\%	& 90\%\\
\cite{lee2017food} & Ultrasonic Doppler Sonar 	& Neck	& ANN	& NA	& 91\%	& 96\%	& 86\%	& 91\%\\

\hline
\multicolumn{4}{l}{\textbf{Food Type Classification}}\\
\hline

\cite{zhang2016diet} & EMG		& Eyes	& RF, LDA	& NA	& 57\%	& NA	& NA	& NA\\
\cite{alshurafa2015recognition} & Piezoelectric	& Neck (solid)	& RF	& NA	& 80\%	& 80\%	& 80\%	& 80\%\\
	&		& Neck (Liquid)	& RF	& NA	& 90\%	& 90\%	& 90\%	& 90\%\\
\cite{lee2017food} & Ultrasonic Doppler Sonar 	& Neck(Chewing)	& ANN	& NA	& 91\%	& 96\%	& 87\%	& \textbf{91\%}\\
	&				& Neck(Swallowing)	& ANN	& NA	& 72\%	& 76\%	& 78\%	& 77\%\\

\hline
\multicolumn{4}{l}{\textbf{Swallow Detection}}\\
\hline
\cite{alshurafa2015recognition} & Piezoelectric	& Neck (solid)	& RF	& NA	& 94\%	& 89\%	& 94\%	& \textbf{91\%}\\
		&	& Neck (Liquid)	& RF	& NA	& 87\%	& 94\%	& 88\%	& \textbf{91\%}\\
\cite{lee2017food} & Ultrasonic Doppler  Sonar 	& Neck	& ANN	& NA	& 78\%	& 75\%	& 76\%	& 76\%\\
\hline 
\end{tabular}
\begin{tablenotes}\footnotesize
\item[*]F1 score is calculated by current authors based on precision and recall values.
\end{tablenotes}
\end{threeparttable}
\end{table*}

\begin{table*}[t]
\caption{Comparison of different evaluation measures for Misc. Multi-Sensor based approaches} 
\label{tab:comparison_multi}
\centering 
\begin{tabular}{p{0.5cm}p{2cm}p{2in}p{1cm}p{1cm}p{1cm}p{1cm}p{1cm}p{1cm}} 
\hline\hline 
\textbf{Ref.}	& \textbf{Sensor} & \textbf{Location}	& \textbf{Algo.}	& \textbf{Env.}	& \textbf{Acc.}	& \textbf{Prec.}	& \textbf{Rec.}	& \textbf{F1} \\ [0.5ex] 
\hline 
\multicolumn{9}{l}{\textbf{Eating Activity Detection}}\\
\hline
\cite{mirtchouk2017recognizing} & Ear, Eyes, Wrist	& Microphone, Motion Sensor, Google Glass	& RF	& L	& 90\%	& 45\%	& 85\%	& 59\%\\
\cite{farooq2016novel} & Eyes, Head	& Piezoelectric, Accelerometer	& SVM	& L	& NA	& NA	& NA	& \textbf{100\%}\\
\cite{anderez2018hierarchical} & Wrist, Waist	& Accelerometer, Tri-axial micro electro mechanical system	& LSTM	& NA	& 99\%	& 99\%	& 99\%	& \textbf{99\%}\\

\cite{bedri2015detecting} & Head, Ear	& Proximity, Gyroscope	& HMM	& L	& NA	& 93\%	& 93\%	& 93\%\\
		& & &				& FL	& NA	& 42\%	& NA	& NA\\
\cite{bedri2015wearable} & Ear, Head	& Proximity, IMU &	HMM	& L	& 95\%	& 93\%	& 96\%	& 95\%\\
\cite{hossain2019human} & Ear 	& Microphone, Accelerometer	& CNN	& NA	& 100\%	& NA	& NA	& NA\\

			& & &					& Misc. 	& 85-92\%	& 31-38\%	& 87-92\%	& NA\\
\cite{farooq2019validation} & Ear, Neck, Wrist& 	Piezoelectric, Accelerometer, Proximity	& ANN	& FL	& NA	& NA	& NA	& 80\%\\
\cite{farooq2017real} & Eyes	& Piezoelectric, Accelerometer	& DT	& NA	& NA	& 93\%	& 94\%	& 93\%\\

\hline
\multicolumn{9}{l}{\textbf{Chewing Detection}}\\
\hline
\cite{bedri2017earbit} & Ear, Neck	& IMU, Microphone, Proximity	& RF	& SCL	& 90\%	& 86\%	& 96\%	& \textbf{91\%}\\
			& & &				& FL	& 93\%	& 81\%	& 79\%	& 81\%\\
\cite{farooq2016detection} & Ear, Wrist, Neck	& Piezoelectric, hand-to-mouth gesture sensor, accelerometer	& LDA	& FL	& 93\%	& 97\%	& 90\%	& 93\%\\
\cite{papapanagiotou2016novel} & Ear	& PPG, Microphone, Accelerometer	& SVM	& SFL	& 94\%	& 79\% &	81\%	& 76\%\\
\cite{zhang2020necksense} & Neck	& IMU, Proximity, Ambient Light	& GBM	& FL	& NA	& 81\%	& 73\%	& 74\%\\
\cite{min2018audio} & Ear	& IMU, Microphone	& RF	& NA	& 73\%	& NA	& NA	& NA\\

\cite{farooq2019validation} & Ear, Neck, Wrist	& Piezoelectric, Accelerometer, Proximity	& ANN	& FL	& NA	& NA	& NA	& 78\%\\

\cite{li2018novel} & Throat	& Microphone, PPG	& SVM	& L	& NA	& 91\%	& NA	& NA\\

\hline
\multicolumn{9}{l}{\textbf{Eating Episode Detection}}\\
\hline

\cite{bedri2017earbit} & Ear, Neck	& IMU, Microphone, Proximity	& HMM	& SCL	& 100\%	& NA	& NA	& NA\\
		& & &					& FL	& 94\%	& NA	& NA	& NA\\
\cite{bedri2020fitbyte} & Eyes	& IMU, Proximity, Mini Spy Camera	& RF	& SFL	& 94\%	& 91\%	94\%	& 93\%\\
\cite{zhang2020necksense} & Neck	& IMU, Proximity, Ambient Light	& DBSCAN	& FL	& NA	& 87\%	& 78\%	& \textbf{77\%}\\
\hline
\multicolumn{9}{l}{\textbf{Food Type Classification}}\\
\hline

\cite{mirtchouk2016automated} & Ear, Wrist, Eyes	& Microphone, 9-axis IMU, Google Glass 9-axis motion sensor	& RF	& L  & 83\%	& NA	& NA	& NA\\

\hline
\textbf{Ref.} & \textbf{Sensors} & 	\textbf{Activity Detected} & 	\textbf{Model}	 & \textbf{Env.} & 	\textbf{Specificity}	 & \textbf{Sensitivity}\\
\hline
\cite{li2018novel} &  Accelerometer, PPG	 & SD	 & SVM	 & NA	 & 60\%	 & 100\%\\

\hline
\multicolumn{6}{l}{\textbf{Weight Estimation}}\\
\hline

\textbf{Ref.}	 &  \multicolumn{2}{l}{\textbf{Sensors}} &  	\multicolumn{2}{l}{\textbf{Location}}	 &  \textbf{Algo.} &  	\textbf{Env.}	 &  \multicolumn{2}{l}{\textbf{Error Metric}}\\
\hline
\cite{yang2019statistical} & \multicolumn{2}{l}{Piezoelectric, Video Camera}	 & \multicolumn{2}{l}{Ear}	 & LR	 & L	 & \multicolumn{2}{l}{RMSE = 250.8}\\
\cite{mirtchouk2016automated}  & \multicolumn{2}{l}{Microphone, 9-axis IMU, Google Glass 9-axis motion sensor}	 & \multicolumn{2}{l}{Ear, Wrist, Eyes}	 & LR	 & L	 & \multicolumn{2}{l}{ARE = 35\%}\\
\hline 
\end{tabular}
\end{table*}

A summary of microphone-based techniques is presented in TABLE \ref{tab:comparison_micro}. The Eating Activity recognition has been performed with maximum 87\% F1 score by applying microphone sensor on wrist using RF classifier in laboratory conditions. Chewing detection has been reported with maximum 88\% F1 Score by placing microphone in ear with CNN classifier in semi free living conditions. Food type detection has yielded 85\% F1 Score with a neck-based microphone applied with DT classifier in laboratory conditions, whereas for solid and liquid detection, the methodology reported 100\% F1 score. 

TABLE \ref{tab:comparison_misc} provides summary of different uni-sensor approaches. It can be seen that an EMG sensor used at eyes, with OCSVM classifier applied in Free living condition has been reported with 95\% F1 Score for eating activity detection. Moreover, chewing detection has yielded an F1 score up to 96\% with piezoelectric sensor applied on glasses in semi-free living conditions using SVM classifier. One approach used Specificity and Sensitivity measures for chewing detection evaluation, with 28\% and 88\% scores respectively.  Eating episodes are detected with maximum 93\% F1 score which is slightly better than IMU-based approaches. Food Type detection has been reported to yield maximum 91\% F1 Score with Ultrasonic Doppler Sonar and ANN Classifier which is greater than other approaches. Swallow detection has achieved maximum 91\% F1 Score using piezoelectric sensor with RF classifier. One approach used specificity and sensitivity measures for swallow detection evaluation with 65\% and 64\% scores respectively. Chewing estimation has yielded a maximum 28\% specificity and 88\% sensitivity; however, there is no other technique which reported same measures. Two more approaches reported chewing estimation, in which, one approach reported 9.7\% Error Rate and the other reported 16\% Percentage Error. 

Summary of multi-sensor based approaches is given in TABLE \ref{tab:comparison_multi}. The highest F1 score was reported to be 100\% for eating activity detection piezoelectric strain sensor and accelerometer. Another approach reported 99\% F1 score with a combination of Accelerometer, Tri-axial micro electro mechanical system applied at wrist and waist with LSTM Classifier.  Chewing detection has achieved a maximum of 93\% F1 score using 3 sensors (Piezoelectric, hand-to-mouth gesture sensor, accelerometer) at ear, wrist and neck with LDA classifier. Eating episode detection has achieved a maximum 93\% score with 3 sensors (IMU, Proximity, Mini Spy Camera) applied at ear, wrist and neck. Food type detection has achieved highest 83\% accuracy using 3 sensors (microphone, 9-axis IMU, Google Glass 9-axis motion sensor) applied in laboratory conditions with RF Classifier, whereas F1 score is not reported in that case. Same approach reported 35\% ARE for weight estimation using Logistic Regression. Weight estimation was reported with 250.8 RMSE in the study \cite{yang2019statistical} which used 2 sensors piezoelectric and video camera) in laboratory conditions using Logistic Regression. The two approaches for weight estimation can’t be compared due to different measures of evaluation.

\section{Analysis and future guidelines}
The performance of uni-sensor and multi-sensor approaches can be compared on the basis of their performance against different tasks. For eating activity detection, the multi-sensor approach \cite{farooq2016novel} outperforms the uni-sensor approaches which works with piezoelectric and accelerometer. For eating episode detection, a PPG sensor based approach \cite{papapanagiotou2016novelconf} performs better than other approaches keeping in view the usage of single sensor. Chewing detection has better performance using piezoelectric sensor based approach \cite{farooq2016segmentation} than any multi-sensor approaches. Swallowing detection has better performance with piezoelectric based single sensor\cite{alshurafa2015recognition}, whereas food classification also yields better performance using single ultra-sonic Doppler sensor\cite{lee2017food} than multi-sensor approaches. 

\begin{table*}[t]
\caption{Comparison of maximum scores achieved by uni-sensor and multi-sensor approaches against different activities} 
\label{tab:comparison_max_scores}
\centering 
\begin{tabular}{p{3cm}p{0.5cm}p{3cm}p{1.5cm}p{0.5cm}p{5cm}p{1.5cm}} 
\hline\hline 
& \textbf{Ref.}	& \textbf{Sensor}	& \textbf{F1 Score}	& \textbf{Ref.}	& \textbf{Sensors}	& \textbf{F1 Score}\\
\hline

Eating Activity Detection & 	\cite{zhang2018free}  & EMG	 & 95\%	 & \cite{farooq2016novel} 	 & Piezoelectric, Accelerometer	 & \textbf{100\%}\\
Swallowing Detection & 	\cite{alshurafa2015recognition}  & Piezoelectric	 & \textbf{91\%} \\
Chewing Detection & 	\cite{farooq2016segmentation}  & Piezoelectric	 & \textbf{96\%}	 & \cite{farooq2016detection}   & Piezoelectric, gesture sensor, accelerometer	 & 93\%\\
Eating Episode Detection & 	\cite{papapanagiotou2016novelconf}  & PPG	 & \textbf{93\%}	 & \cite{bedri2020fitbyte} 	 & IMU, Proximity, Mini Spy Camera	 & \textbf{93\%}\\
Food Type Classification & 	\cite{lee2017food}  & Ultrasonic Doppler Sonar	 & \textbf{91\%}	 & \cite{mirtchouk2016automated} &  Microphone,9-axis IMU, Google Glass	 & 83\% (Acc.)\\
\hline 
\end{tabular}
\end{table*}

\subsection{Limitations}
\begin{enumerate}
\item IMU-based approaches can only detect monitions like wrist motions, finger movements, head movement and jaw bone movement. The approaches are limited to eating activity detection, or eating episode detection. 
\item Some approaches utilize head movement to recognize chewing but there is no approach which follows up with swallowing detection due to the limitation of sensors. Therefore, approaches based on IMU sensors only, cannot provide comprehensive solution.
\item Microphone based approaches have low performance for all tasks mainly due to the noise related vulnerabilities in free living environment. The maximum chewing detection F1 score is 88\% compared to 96\% F1 score which is the highest in this domain.
\item Ultrasonic Doppler Sonar based approach has best F1 score for food type classification but its effect of continuous signal emission on human body is unknown. 
\item Multi-sensor based approaches out-performed other approaches for eating activity only. For other tasks, the F1 score are 3\%-5\% lower as compared to uni-sensor based approaches. Additionally, these approaches require multiple sensors attachment on body resulting in obtrusive nature.
\item Only few approaches targeted food type classification, whereas only one approach provided more than 90\% F1 score. 
\item A small number of approaches targeted estimation of food mass and duration using Lin. R model or Log. R model. However, usage of dissimilar error metrics limits the comparison of these approaches. 
\item Most approaches don’t take chewing and swallowing as basic features of eating detection, as these are the foundation of a reliable food monitoring system.
\item The food intake monitoring domain has not utilized recent feature selection approaches (semi-supervised feature selection methods and causal feature selection methods) which are not only efficient in computation, but also provide better feature subsets for predictions.
\item There has been a considerable work done in the area of time-series classification in machine learning. Food intake monitoring has not taken advantage of these recent algorithms as main emphasis has remained on usage of conventional classification algorithms. Moreover, ensemble learning provides more robust solutions to classification performance, which are not considered in food intake monitoring domain.
\item There is no approach comprehensively covering all tasks of automatic food monitoring systems.
\end{enumerate}

\subsection{Future Directions}
In this section we setup directions for future research in the area of food activity monitoring systems using wearable sensors. We suggest directions particularly for the selection of sensors, feature selection methods and classification algorithms based upon the limitations in literature under study and some recent advances particularly in feature selection and learning algorithms.
\subsubsection{Selection of Sensors}
We have noticed from the review of previous approaches that there is no single approach which deals with all of the food intake monitoring system tasks. On the other hand, these tasks were achieved up to an acceptable mark by some researchers using either a single sensor or multiple sensors. 

Chewing detection is the foundation of automatic food monitoring systems. An eye-glasses mounted piezoelectric sensor approach \cite{farooq2016segmentation} has reported 96\% F1 score for chewing detection. In another approach \cite{alshurafa2015recognition} the swallowing detection F1 score was reported as 91\% using piezoelectric sensor which is the best swallowing detection score in all approaches. However, the location of sensor in the later approach is neck instead of eyes. The same approach yielded 90\% and 80\% F1 score for Food Type classification for liquid and solid foods respectively. Since eating episodes are temporal activities based upon the underlying chewing and swallowing activities, therefore, the sensor has strong chances to perform better for eating episode detection as well. Conclusively, piezoelectric sensor has the ability to provide a comprehensive solution in this context. However, location identification for all activities remains a question. 

In chewing detection perspective, a multi-sensor based approach \cite{farooq2016detection} has achieved up to 93\% F1 score which includes Piezoelectric, hand-to-mouth gesture sensor and accelerometer. Piezoelectric sensor on the other hand has been reported with best swallowing detection score \cite{alshurafa2015recognition}. Secondly, piezoelectric sensor provides 91\% and 80\% food classification F1 Score \cite{alshurafa2015recognition} for liquid and solid foods respectively. Apart from this, accelerometer has reported best eating activity recognition results \cite{anderez2018hierarchical}. The research work reported in \cite{farooq2016detection} can be utilized as the foundation of a more comprehensive system by integrating the other three approaches mentioned here. The challenge in this case will be to retain the swallowing detection, activity detection and food classification scores in integrated fashion using same classifier (LDA). Another key challenge will be to make it unobtrusive, since it has 3 sensors at different locations on the body. 

An ultrasonic Doppler Sonar based approach \cite{lee2017food} has successfully achieved 91\% chewing detection F1 score, whereas its food type classification results are better than all uni-sensor and multi-sensor approaches. However, the swallowing detection F1 score is 75\% which is lower than many approaches. Moreover, the effect of continuous beam emission on human body needs to be studied in this regard. However, since there is no other approach with this sensor, a detailed investigation can be done to make a comprehensive solution.

\subsubsection{Feature Selection Methods}
Feature selection plays an important role in food intake monitoring systems, as the signal data is usually high dimensional and not all of the features are relevant proposed in \cite{bi2018auracle} used only 40 features out of 700 features, \cite{shin2019accurate} used only 23 out of 900 extracted features and \cite{kalantarian2015audio} used only 13 features out of 6555 features.

From the literature review in this study, it can be seen that most researchers focused on the conventional feature selection methods \cite{bi2018auracle,farooq2018accelerometer} which are prone to certain disadvantages, whereas the feature selection domain has evolved in previous years. These techniques are divided into unsupervised, supervised and semi-supervised methods \cite{chandrashekar2014survey}. The unsupervised methods are difficult to adapt due to non-availability of class labels. Supervised feature selection algorithms select relevant features where labeled data is available \cite{chandrashekar2014survey,jiang2019probabilistic,he2015robust}, whereas semi-supervised algorithms allow feature selection by handling both labeled and unlabeled data \cite{zhao2007semi,liu2013efficient,xu2010discriminative}, thereby advantageous over supervised algorithms. Recently, a new approach for semi-supervised feature selection handled the issue of noisy unlabeled samples \cite{chang2014convex}, however it doesn't address the problem of local relationships, which is an important issue in case of high dimension features with small number of label instances. Recently \citeauthor{jiang2019joint} \cite{jiang2019joint} proposed an approach called Joint Semi-supervised Feature Selection (JSFS) to handle this issue. This algorithm is combined with a Bayesian approach in order to prune relevant features and at the same time learns a classifier. It doesn't require the number of features in advance, as well as doesn't require additional classifier while it shows robustness towards noise in unlabeled data. 

Another important concept in feature selection is to exploit the relationships between features and class labels by using causal feature selection techniques \cite{yu2018unified,wu2020tolerant}. These methods are optimal when compared to non-causal feature selection methods \cite{tsamardinos2003towards}. The concept of Markov Boundaries was introduced by \citeauthor{koller1996toward} \cite{koller1996toward} in causal feature selection which is to define a boundary for a class attribute constructing the relationships between the class attribute and features. It leads to more robust predictions compared to the conventional feature selection methods. Some recent studies in this regard include incremental association MB (IAMB) including some of its other variants \cite{tsamardinos2003towards}, Min–max MB (MMMB) \cite{tsamardinos2003time} and e HITION-MB \cite{aliferis2003hiton,aliferis2010local}. In a more recent study \citeauthor{wu2019accurate} \cite{wu2019accurate} have proposed cross-check and complement MB discovery (CCMB) algorithm and pipeline machine-based CCMB (PM-CCMB) algorithm. These algorithms have been proven to yield better accuracy and time efficiency in MB discovery. In another study \citeauthor{wu2020multi} \cite{wu2020multi} proposed Markov blanket based multi-label causal feature selection method which works in a multi-label data environment to select predictive features by using causal relationships.  We believe that food intake monitoring systems can take advantage from these recent advances in feature selection.

\subsubsection{Selection of Learning Algorithm}
We also notice that many research studies in food activity monitoring have used conventional classification algorithms in the learning phase, and has not utilized advanced forms of these algorithms.

Recently, ensemble learning has caught attention which combine multiple algorithms (weak learning algorithms) in order to create a single prediction model with the target to achieve better results in terms of prediction with low variance and bias. In this context, Probabilistic Classification Vector Machines (PCVM) were introduced by \citeauthor{chen2009probabilistic} \cite{chen2009probabilistic} who proposed to take benefits from a signed and truncated Gaussian Prior to generate sparsity. This approach achieves better results that Support Vector Machines (SVM-Hard/Soft) and Relevance Vector Machines\cite{tipping2000relevance}. Since the PCVM only works for binary classes, a multi-class PCVM is also introduced by \citeauthor{lyu2019multiclass} \cite{lyu2019multiclass}. These techniques suffer from sensitivity to local minima due to the usage of Expectation Maximization Algorithm, and also sensitive to large data sets. An improved version has been proposed by \citeauthor{chen2013efficient} \cite{chen2013efficient} which is termed as Efficient Probabilistic Classification Vector Machine (EPCVM). The new algorithm takes advantages from Laplace Approximation and Expectation Propagation which results in tackling the issues discussed before. These techniques can be used to achieve better prediction results in classification phase of food intake monitoring systems.

Since the classification problem in the food intake monitoring is a sequential learning process, there appears to be a lack of usage recent approaches in this aspect. For instance, recently "learning in the model space" has caught attention in the research community, in which important data from the original data set is captured through models and algorithms are applied on those models, rather than the original data. In this aspect, the proposed method by \citeauthor{chen2013model} \cite{chen2013model} has the ability to handle time series classification problem in a computationally efficient manner with high classification accuracy compared to other similar methods where learning is performed in model space. Apart from this, the proposed methods can handle long time series data efficiently. \citeauthor{gong2018multiobjective} \cite{gong2018multiobjective} is important for time series classification, so they have proposed a learning algorithm called multi-objective model-metric (MOMM) which takes benefits from time-series and label information both whereas the data is classified in the model space. This algorithm works better as compared to other algorithms like DTW, MESN and Fisher. Another approach in this context is proposed by \citeauthor{li2019short} \cite{li2019short} which improves over Linear dynamical system (LDS) for short sequence classification. \citeauthor{chen2013learning} \cite{chen2013learning} proposed another framework in which the model space is explored instead of signal space to perform classification in fault diagnosis domain. In this approach multiple model spaces are fitted with the help of series of signal segments which are selected using a sliding window.  Measuring the distance between bite sequences is an important sub-topic in food intake monitoring. Studies suggests that Dynamic Time Wrapping (DTW) has shown good results compared with other approaches in video, images and audio signals. However, this method is sensitive to noise. An improved method is designed by \citeauthor{gong2018sequential}  \cite{gong2018sequential} termed as Dynamic State Wrapping (DSW) which is coupled with 1-Nearest Neighbour (1NN) classifier. It has shown more robustness and fits better in case of long sequences compared with DTW. We believe that using these approaches in food intake monitoring can achieve better results in terms of classification accuracy and computation time.

\section{CONCLUSION}
Automatic food intake monitoring has become a vibrant domain in recent past, as manual food monitoring systems are difficult to adapt due the biased-reporting and manual record keeping. To tackle this challenge, a variety of wearable sensors have been used by the researchers which majorly separated eating out of other daily activities initially. Afterwards a considerable effort has been put on chewing, swallowing and food type detection. Research community has also worked in the area of food mass estimation and intake duration estimation by deploying wearable sensors on the human body. In this research study, we have conducted a critical analysis of automatic food activity monitoring systems which were designed in the last 5 years using wearable sensors. The study concludes that there is no system which can provide all functionalities of a food activity monitoring system. We have provided interesting facts based upon the results reported by these studies, demonstrating the comparison between the results of one sensor with other sensors under different conditions using various parameters. We have also setup future directions which can assist the upcoming research in this domain based upon the limitations and latest literature across feature selection and learning algorithm domains.

 \clearpage 
\appendices
\section{}
\begin{table}[!htbp]
\caption{List of Abbreviations} 
\centering 
\begin{tabular}{p{1cm}p{7cm}}
\hline\hline
\multicolumn{2}{l}{\textbf{Classifiers and Algorithms}}\\
\hline

CBPA & Chewing Band Power Algorithm\\
LCA	& Level Crossing Algorithm\\
JSC	& Jaccard Similarity Coefficient\\ 
DNN	& Deep Neural Network\\
GBM & Gradient Boosting Machine	\\
HMM	& Hidden Markov Models\\
CNN	& Convolutional Neural Network	\\
Lin. R	& Linear Regression\\
MCC	& Mean Crossing Counts\\
WPD	& Window Peak Count\\
WSVM & Weighted Support Vector Machine\\
FFS	& Forward Feature Selection\\
MEC	& Motion Energy Calculation\\
MLP	& Multi-Layer Perceptron\\
FIRF & Finite Impulse Response Filter\\
RNN	& Recurrent Neural Network\\
NB	& Naive Bayes\\
AMAE	& Average Mean Absolute Error\\
DT	& Decision Tree\\
LPFA	& Low Pass Filtering Algorithm\\
RBMs	& Restricted Boltzmans Machines\\
SVM	& Support Vector Machines\\
LSTM	& Long Short Term Memory Network\\
RF	& Random Forest\\
KNN	& K-Nearest Neighbours\\
Log. R	& Logistic Regression\\
mRMR	& Min. Redundancy and Max. Relevance\\
ANN	& Artificial Neural Network\\
FFS	& Forward Feature Selection\\
R FEA	& Recursive Free Elimination Algorithm\\
FSM	& Forward Selection Method\\
HSW	& Hanning Sliding Window\\
CFSSE	& Correlation Feature Selection Subset Evaluator\\
ocSVM	& One-Class Support Vector Machines\\
MSEA	& Maximum Sound/Signal Energy Algorithm\\
DBSCAN	& Density Based Spatial Clustering of Applications with Noise\\
LPSSA	& Longest Period Sub Sequence Algorithm\\
EMA	& Exponentially Weighted Moving Average\\
PBPF	& Prominence Based Peak Finding Algorithm\\
SFFS	& Sequential Forward Floating Selection Algorithm\\

\hline
\multicolumn{2}{l}{\textbf{Miscellaneous (Sensors, Metrics etc)}}\\
\hline

AME	& Absolute Mean Error\\
IMU	& Inertial Management Unit\\
MSSE	& Mean Square Standard Error\\
EMG	& Electromyographic\\
RME	& Root Mean Error\\
PPG	& Photoplethysmography\\
LOSO	& Leave One Subject Out\\
CD	& Chewing Detection\\
LOPO	& Leave One Person Out\\
FTD	& Food Type Detection\\
L	& Laboratory\\
EED	& Eating Episode Detection\\
FL	& Free Living\\
FEE	& Food Energy Estimation\\
SCL	& Semi Controlled Laboratory\\
CE	& Chewing Estimation\\
SFL	& Semi Free Living\\
NA	& Not Available\\
SE	& Swallowing Estimation\\
EDE	& Eating Duration Estimation\\
SD	& Swallowing Detection\\
EAD	& Eating Activity Detection\\

\hline 
\end{tabular}
\end{table}

\ifCLASSOPTIONcaptionsoff
  \newpage
\fi
\printbibliography
\begin{IEEEbiography}[{\includegraphics[width=1in,height=1.25in,clip,keepaspectratio]{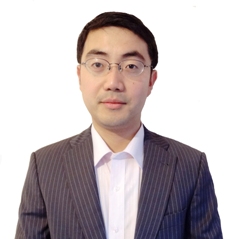}}]{Huanhuan Chen}
(M’09–SM’16) received the B.Sc degree from the University of Science and Technology of China (USTC), Hefei, China, in 2004 and the Ph.D degree in computer science from the University of Birmingham, Birmingham, UK, in 2008. He is currently a Full Professor in the School of Computer Science and Technology, USTC. His research interests include neural networks, Bayesian inference and evolutionary computation. Dr. Chen received the 2015 International Neural Network Society Young Investigator Award, the 2012 IEEE Computational Intelligence Society Outstanding Ph.D. Dissertation Award, the IEEE TRANSACTIONS ON NEURAL NETWORKS Outstanding Paper Award (bestowed in 2011 and only one paper in 2009), and the 2009 British Computer Society Distinguished Dissertations Award. He is an Associate Editor of the IEEE TRANSACTIONS ON NEURAL NETWORKS AND LEARNING SYSTEMS, and the IEEE TRANSACTIONS ON EMERGING TOPICS IN COMPUTATIONAL INTELLIGENCE.
\end{IEEEbiography}
\begin{IEEEbiography}[{\includegraphics[width=1in,height=1.25in,clip,keepaspectratio]{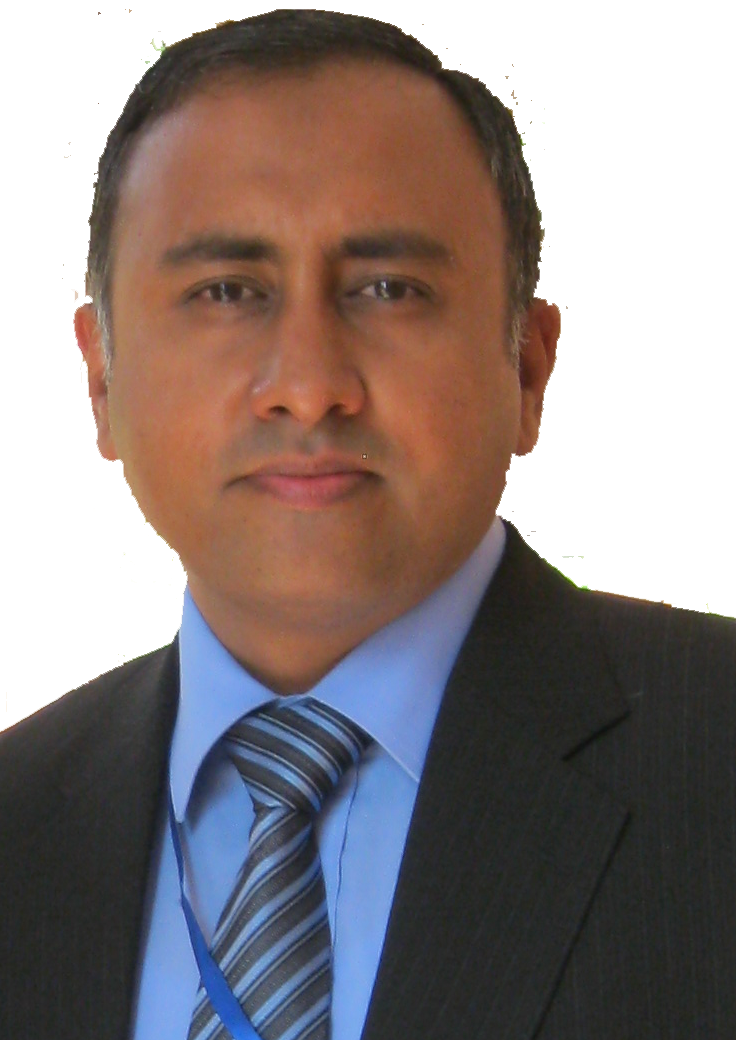}}]{Muhammad Usman}
received his B.Sc degree from International Islamic University, Islamabad, Pakistan in 2006, Master's degree from SZABIST, Islamabad, Pakistan in 2017. He is currently pursuing the Ph.D degree in Computer Science at the School of Computer Science and Technology, USTC, Hefei, China. He has been working at Pakistan Scientific and Technological Information Center, MoST, Pakistan since 2010 as Database Administrator. His research interests include Machine Learning, Data Mining, Data Warehousing and Knowledge Discovery. He is currently working in Probabilistic Classification Algorithms and Wearable Sensors. Previously, he has worked in application of data mining techniques in data warehouses for knowledge discovery without the requirement of domain knowledge. He has published his work in at international conferences, journals and books.
\end{IEEEbiography}
\end{document}